\documentclass[twocolumn]{autart}    

\usepackage{graphicx}          
\usepackage{amssymb}
\usepackage{multirow}
\usepackage{makecell}
\usepackage{booktabs}
\usepackage{graphicx}
\usepackage{lineno}
\usepackage[cmintegrals]{newtxmath}
\usepackage{bm}
\usepackage{amsmath}
\usepackage{enumerate}
\usepackage{color}
\usepackage{graphics}
\usepackage{multicol}
\newcommand{\ds}{\displaystyle}
\newtheorem{theorem}{Theorem}[section]

\newtheorem{definition}[theorem]{Definition}
\newtheorem{problem}[theorem]{Problem}

\newtheorem{proposition}[theorem]{Proposition}
\newtheorem{lemma}[theorem]{Lemma}
\newtheorem{example}[theorem]{Example}
\newtheorem{procedure}[theorem]{Procedure}

\begin{document}

\begin{frontmatter}

\title{\large A Critical Escape Probability Formulation for Enhancing 
the Transient Stability of Power Systems with System Parameter Design} 

\thanks[footnoteinfo]{
 Corresponding author: Kaihua Xi.}

\author[SDU]{Xian Wu}\ead{xianwu@mail.sdu.edu.cn
},    
\author[SDU]{Kaihua Xi}\ead{kxi@sdu.edu.cn},               
\author[SDU]{Aijie Cheng}\ead{aijie@sdu.edu.cn},  
\author[SDU2]{Chenghui Zhang}\ead{zchui@sdu.edu.cn}, 
\author[TUD,LEIDEN]{Hai Xiang Lin}\ead{H.X.Lin@tudelft.nl}

\address[SDU]{School of Mathematics, Shandong University, Jinan, 250100, P. R. China }  
\address[SDU2]{School of Control Science and Engineering, Shandong University, Jinan, 250061, P. R. China}     
\address[TUD]{Delft Institute of Applied Mathematics, Delft University of Technology, Delft, 2628 CD, The Netherlands}
\address[LEIDEN]{Institute of Environmental Sciences (CML), Leiden University, Leiden 2333 CC, The Netherlands}            


\begin{abstract}
For the enhancement of the transient stability of power systems, the key
is to define a quantitative optimization formulation
with system parameters as decision variables. 
In this paper, we model the disturbances by Gaussian noise and 
define a metric named \emph{Critical Escape Probability} (CREP) based on the 
invariant probability measure of a linearised stochastic processes.
CREP characterizes the probability of the state escaping from a critical set. CREP involves all the system parameters and reflects the size of the basin of attraction of the nonlinear systems. An optimization framework that minimizes CREP with the system parameters
as decision variablesis is presented. Simulations show that the mean first hitting time when the state hits the boundary 
of the critical set, that is often used to describe the stability of nonlinear systems, is dramatically 
increased by minimizing CREP. This indicates that the transient stability of the system
is effectively enhanced. It also shown that suppressing the state fluctuations only is insufficient 
for enhancing the transient stability. In addition, the famous Braess' paradox which also exists in power systems is revisited. Surprisingly, it turned out that the paradoxes identified by the traditional metric may not exist according to CREP.
This new metric opens a new avenue for the transient stability analysis of future power systems integrated with large amounts of renewable energy.
\end{abstract}

\end{frontmatter}

\section{INTRODUCTION}
In the synchronous state of a power system, the frequencies of all synchronous machine must be at or near the nominal frequency (eg, 50 Hz or 60 Hz). The frequency is the derivative of the rotational phase 
angle and is equal to the rotational speed of the synchronous machine expressed in units of rad/s. 
Synchronization of the frequency is essential for the proper functioning of a power system. Severe interference can cause desynchronization, which can lead to widespread power outages. Current energy systems are moving towards more distributed generation by renewables, which tend to be inherently more uncertain and low inertial, posing an even greater threat to synchrony. 
\par 
Synchronization stability, which is also called \emph{the transient stability} in power engineering, 
is the ability to maintain the synchronization when subjected to disturbances \cite{Kundur1994}. 
In this paper, we refer to synchronization stability and transient stability equivalently. 
The synchronous state
and its stability are determined by the system parameters which include the 
power generation and loads, the inertia and the damping of the synchronous machines, the capacity 
of lines and the network topology.  For deterministic systems, significant 
insights on the role of these parameters have been obtained from investigations on the existence condition of 
a synchronous state \cite{Dorfler20141539,DorflerCriticalcoupling}, the linear or nonlinear stability \cite{pecora}, the synchronization coherence \cite{FAZLYAB2017181} and the basin of attraction \cite{menck,sizeofbasin}.
The system parameters may be assigned to optimize
the synchrony, which can be obtained by load frequency control, the placement of virtual 
inertia, configuration of the damping coefficient, deletion or addition of lines or by changing 
the line capacities. In particular, regarding the transient stability, the local convergence to the synchronous state or the basin of attraction of the synchronous state are investigated \cite{hirsch1,zab2}.
\par
In practice, the synchronous state of the power system is a set point for control,
in which control actions are taken to let the state converge to this set point after disturbances.
Thus, with frequently occurring disturbances, e.g., the uncertainties from wind energy and power demand
and unpredictable fault in power generation, 
the frequency, and the phase usually fluctuate around the synchronous state. 
If both the fluctuations of the frequency and phase difference between the synchronous machines are so 
large that the state of the system cannot return to the synchronous state, then the synchronization is lost.
Hence, the risk of losing synchronization is actually 
determined by two factors, i.e., the size of the basin of attraction of the synchronous state and 
the fluctuation of the state caused by the disturbances. To increase the 
transient stability, it is important to find such a synchronous state that 
has a large basin of attraction and around which the fluctuation of the state is also small. 
It is insufficient 
to analyze the transient stability in a deterministic system without considering the state fluctuations caused 
by the disturbances. 
\par 
Regarding the state fluctuations, 
various investigations have been made to learn the impacts of the system parameters, from 
which insights have been obtained on the propagation of the disturbances and the parameter assignment for suppressing the fluctuations. With perturbations added to the system parameters, the disturbance
arrival time is estimated
in \cite{topological-spreading}. The
amplitude of perturbation responses of the nodes is used to
study the emergent complex response patterns across the network in \cite{Zha19}. 
By modelling the disturbances as inputs to an associated linearized system,
the fluctuations are evaluated by the $\mathcal{H}_2$ norm of the input-output linear system \cite{H2norm_another_form,optimal_inertia_placement,H2norm}. By minimizing it, the fluctuations can also be effectively suppressed by system parameter assignment, such 
as the optimal placement of virtual inertia \cite{optimal_inertia_placement}.
To precisely characterize
the fluctuations, the variances of the frequency at each node and the phase difference 
at each line in the invariant probability distribution are investigated  \cite{WANG2023110884, WuXian2} with the disturbance modelled by Gaussian noise.
It is found that the impacts of the disturbances at the nodes can be described by 
\emph{the Superposition Principle} \cite{WuXian2}.
With assumption of uniform 
disturbance-damping ratios among the nodes, explicit formulas 
of the variance have been deduced. From these formulas it is found
that the fluctuations are related to the cycle space of graphs \cite{WANG2023110884}. 
In control theory, the robust control method is applicable to suppress 
the fluctuations by controlling the power generation in load frequency control. 
However, for enhancing the transient stability, it is insufficient to suppress the 
fluctuations only because the stability also depends on the basin of attraction.
\par 
For enhancing the transient stability, the most difficult problem is 
to define a quantitative optimization formulation with the system parameters as the decision variables.
 The mean of the first hitting time when the state hits the boundary of the basin of attraction
is often used to study the survival
time of a system, which is also used to study the stability
of non-linear systems\cite{survivalanalysis,stochasticstability}. 
The longer is the mean first hitting time, the more stable is the system under stochastic disturbances. 
Both of the basin of attraction and the severity of the 
state fluctuations are involve into this value,
which makes it a potential metric for the transient stability. 
However, it can hardly be maximized directly
because it is difficult to get the probability distribution of the first hitting time and the 
boundary of the basin of attraction.  
For coupled phase oscillators, the probability that the state exits a secure domain,
which also involves the basin of attraction and the state fluctuations,
is investigated in order to enhance the synchronization stability of the system 
in \cite{WuXian}. However, the dynamics of the frequencies at the nodes are not considered 
in that system. 
\par 
In this paper, for power systems with stochastic disturbances, we model the disturbances by Gaussian noises and 
focus on the invariant probability distribution of the frequency and the phase difference in a linearized 
stochastic process. We define a metric named \emph{Critical Escape Probability}(CREP),
 which describes the probability of the state escaping from a critical set,
 to assess the transient stability. It is related to the mean first hitting time of the state to the boundary of 
 the critical set, i.e., the smaller is CREP, the longer is the mean first hitting time.
We analyze the trends 
of CREP as the system parameters change and its relationship to the size of the basin of attraction. 
 In addition, we revisit the famous Braess' paradox \cite{Witthaut2012,Coletta2016} with CREP. It is found 
 this paradox can also be identified by CREP. In particular, it is surprisingly found
 that adding a new line may lead to increasing the stability under CREP while
 decreasing the stability under the other existing metrics.  
 This is because  the influences of all the system parameters are included into CREP while in the other metrics, e.g.,
 the linear stability measured by the spectrum of the Jacobi matrix and the order parameter defined by Kuramoto to study the level of the synchronization, not all system parameter's influences are fully considered.  
We formulate an optimization framework that minimizes CREP with the system parameters as 
decision variables. The mean first hitting time is used to verify the performance 
of CREP on identifying the Braess' paradox and the optimization framework 
on enhancing the transient stability.  
The optimization framework can be applied in optimal power flow calculation, the placement
of virtual inertia, tuning the gain for droop control and the design of the network topology. 
It also provides a new avenue to the stability analysis of the complex system in which the synchronization plays
an important role on the proper function of the system \cite{Dorfler20141539}.  
\par 
The contributions of this paper include:
\begin{enumerate}[(1)]
\item CREP formulation for assessing the transient stability, which involves
the roles of all the system parameters and can be minimized to enhance the transient stability;
\item An optimization framework that minimizes CREP,  by which the system parameters 
can be optimally configured to increase the first hitting time, thus enhance the transient stability
of the system under stochastic disturbances;
\item A new finding on the identification of the Braess' paradox by CREP. 
\end{enumerate}
\par 
This paper is organized as follows. We formulate the problem by introducing the mathematical model 
of power system and the concept of the mean first hitting time in Section \ref{Section:Model}. 
The invariant probability distribution of a linear stochastic process and the definition of CREP are described in 
Section \ref{Section:invariantprobability} and the optimization 
framework for improving the transient stability is presented in Section \ref{Section:Optimization Framework}.
We analyze the dependence of CREP on the system parameters and evaluate performance of the proposed optimization framework on improving the transient stability through case studies in Section \ref{Section:Case study} and conclude
with remarks in Section \ref{Section:conclusion}.

\section{Problem formulation}\label{Section:Model}
In this section, we present the scientific problem of this paper with the introduction of the model and the 
mean first hitting time of a stochastic process. 
\subsection{The model}
The network of the power system can be modelled by a graph $\mathcal{G}=(\mathcal{V},\mathcal{E})$ 
with $n$ nodes in set $\mathcal{V}$ and $m$ lines in  set $\mathcal{E}\subset \mathcal{V}\times\mathcal{V}$, where a node denotes a bus and a line denotes a transmission line connecting two buses. We focus on the transmission network 
and assume the lines are lossless. The dynamics of the power system are described by the swing equations
\cite{zab2,menck2,hirsch1}
\begin{subequations}\label{Nonlinear}
\begin{align}
\dot{\delta_i}(t)&=\omega_i(t),\\
m_i\dot{\omega_i}(t)&=P_i-d_i\omega_i(t)-\sum\limits_{j=1}^n l_{i,j}\sin (\delta_i(t)-\delta_j(t)),
\end{align}
\end{subequations}
where $\delta_i(t)$ and $\omega_i(t)$ denote the phase angle and the frequency deviation from 
the nominal frequency of the synchronous machine at node $i$; $m_i>0$ describes the inertia of the synchronous generators; $d_i>0$ represents the damping coefficient with droop control; $P_i$ denotes power generation if $P_i>0$ and denotes power load otherwise; $l_{i,j}=\hat{b}_{i,j}V_iV_j$ is the effective susceptance, where $\hat{b}_{i,j}$ is the susceptance of the line $(i,j)$ and $V_i$ is the voltage. In this paper, $l_{i,j}$ is also referred as the \emph{line capacity}. We assume that
the voltage at each node is a constant because the dynamics of the voltage 
and that of the frequency can be decoupled in the stability analysis (\ref{Nonlinear}).
It is assumed that the graph
is connected, thus it holds $m>n-1$. 
\par
When the power generations and loads are time invariant, the frequencies at the nodes synchronize at an equilibrium state, called \emph{the synchronous state} that satisfies, for $i=1,2,\cdots,n$,
\begin{equation*}
    \omega_i(t)=\omega_{syn}~ ~\text{and}~~\omega_{syn}-\frac{\sum_{i=1}^n P_i}{\sum_{i=1}^n d_i}=0. 
\end{equation*}
Without loss of generality, we assume $\sum_{i=1}^n P_i=0$, which means that the power generations and loads are balanced. In practice, this balance is achieved by secondary frequency control \cite{PIAC3}. 
Hence, at the synchronous state, it holds that $\omega_{syn}=0$, and 
the phases $\delta^*_i$ at the nodes satisfies,  
\begin{equation}\label{syn state}
  \begin{aligned}
    P_i-\sum\limits_{j=1}^n l_{i,j}\sin(\delta_{i}^*-\delta_{j}^*)=0.
\end{aligned}
\end{equation}
Clearly, the existence of a synchronous state depends on the topology
structure, the distribution of power generations and loads
at each node and the line capacities \cite{Dorfler20141539,DorflerCriticalcoupling}.
Denote the synchronous state by $((\bm \delta^*)^\top,\bm 0)^\top\in\mathbb{R}^{2n}$
with $\bm \delta^*=\text{col}(\delta^*_i)\in\mathbb{R}^n$. 
For practical reasons,  
we restrict our attention to the synchronous state with the phase in the 
following domain 
\begin{align}
\Theta_\delta &=\{\bm\delta\in\mathbb{R}^{n}\big||\delta_{i}-\delta_j|<\pi/2,\forall (i,j)\in\mathcal{E}\}. \label{Theta_delta}
\end{align}
It has been proven there exists at most one synchronous state in this domain and when it exists,
it is asymptotically stable \cite{skar_uniqueness_equilibrium}. The stability region of the synchronous state has been analyzed by \cite{hirsch1} and
independently by \cite{zab2}. 
\par 
In real networks, the state of the power system always fluctuates around the synchronous state due to various
disturbances. When the fluctuations are very large, the state
may exit the stability region of the synchronous state and become instable.  Desynchronization
means that both the fluctuations of the frequency and
the phase angle difference are so large that the system
cannot return to the synchronous state. 
The fluctuations 
depend on many factors, which include the line capacity,
the inertia and damping of the synchronous machines, the network 
topology and the strength of disturbances. The source of
the disturbances are also various, e.g., the renewable
power generation, fault of the devices in the network,
etc.  We focus on the following problem.
\begin{problem}\label{problem}
How to improve the transient stability of the system under stochastic disturbances by 
changing the system parameters? 
\end{problem}
\par 
To address this problem, we model the disturbance by Gaussian noise and focus on the following stochastic process, 
\begin{subequations}\label{stochasticsystem}
    \begin{align}
    \text{d}\delta_i(t)&=\omega_i(t)\text{d}t,\\
    m_i\text{d}\omega_i(t)&=(P_i-d_i\omega_i(t)-\sum\limits_{j=1}^n l_{i,j}\sin (\delta_{ij}(t)))\text{d}t+b_i \text{d}v_i(t),
    \end{align}
\end{subequations}
where $\delta_{ij}(t)=\delta_i(t)-\delta_j(t)$, $b_i$ is used to describe the strength of the noise, $v_i(t)$
is a Brownian motion process, which has increments with a Gaussian probability distribution. Here, we have assumed for any two distinct nodes $i$ and $j$ the stochastic process
$v_i(t)$ and $v_j(t)$ are independent.  This is reasonable
because the locations of the renewable power generators with serious power generation uncertainties are usually far from
each other.

\par
Denote $e_k=(i,j)\in\mathcal{E}$ for $k=1,\cdots,m$. 
To obtain the information of the frequency and the phase difference, 
we define the output of the system (\ref{Nonlinear}) as the frequencies 
at the nodes and the phase differences in the lines as follows,
\begin{align*}
\bm y(t)=\bm C\bm x(t),
\bm x(t)=\begin{bmatrix}
\bm \delta(t)\\
\bm \omega(t)\\
\end{bmatrix},
\bm y(t)=
\begin{bmatrix}
\bm y_\delta(t)\\
\bm y_\omega(t)
\end{bmatrix},
\bm C=
\begin{bmatrix}
\widetilde{\bm C}^\top&\bm 0\\
\bm 0&\bm I_n
\end{bmatrix},
\end{align*}
where $\bm y\in\mathbb{R}^{n+m}$, $\bm C\in\mathbb{R}^{(m+n)\times 2n}$, $\bm x\in\mathbb{R}^{2n}$,  $\bm\delta=\text{col}(\delta_i)\in\mathbb{R}^n,\, \bm\omega=\text{col}(\omega_i)\in\mathbb{R}^n$, $\bm y_\delta=\text{col}(y_{\delta_k})\in\mathbb{R}^{m}$ with $y_{\delta_k}=\delta_i-\delta_j$ for $k=1,\cdots,m$, $\bm y_\omega=\bm \omega\in\mathbb{R}^{n}$, $\widetilde{\bm C}=(C_{ik})\in\mathbb{R}^{n\times m}$ is the incidence matrix of graph $\mathcal{G}$ such that
\begin{eqnarray}\label{IncidenceDef}
    C_{i,k}
& = & \left\{
      \begin{array}{rl}
         1, & \text{if node $i$ is the beginning of line $e_k$},\\
        -1, & \text{if node $i$ is the end of line $e_k$},\\
         0, & \text{otherwise},
      \end{array}
      \right.
\end{eqnarray}
where the direction of line $e_k$ is specified arbitrarily without influence on the study below, $\bm I_n\in\mathbb{R}^{n\times n}$ is an identity matrix.
Note that the first $m$ elements of $\bm y$ are the phase differences in the $m$ lines and 
the next $n$ elements are the frequencies at the $n$ nodes. At the synchronous state $\bm x^*=((\bm\delta^*)^\top,\bm 0)^\top$, the output becomes
\begin{align}\label{output}
\bm y^*=\bm C\bm x^*=\big((\widetilde{\bm C}^\top\bm\delta^*)^\top,\bm 0\big)^\top.
\end{align}
To address Problem \ref{problem}, a metric that fully 
reflect the transient stability and can be minimized or maximized as an objective function 
in an optimization problem is sought. The mean first hitting time  that is often used to describe the stability of a nonlinear system is introduced below.
\begin{definition}
Consider a stochastic process $\{\bm x(t)\in\mathbb{X},~t\in\mathbb{T}\}$ with initial state $\bm x(0)=\bm x_0$ and a boundary set $\mathbb{B}$ of set $\mathbb{A}$, in which $\mathbb{A}\subset\mathbb{X}$ and $\mathbb{T}=[0,+\infty)$. Assume that the initial value $x_0$ of the process lies inside $\mathbb{A}$ but outside $\mathbb{B}$, then the first hitting time is defined by the random variable $t_e:\Omega\rightarrow\mathbb{R}\cup\{+\infty\}$, 
\begin{align*}
t_e=
\begin{cases}
\inf_{t\in \mathbb{T}} x(t)\in \mathbb{B}, &\text{if such a } t\in\mathbb{R}~~\text{exists},\\
+\infty,  &\text{else},
\end{cases}
\end{align*}
where $t_e$ is the first time when the sample path of the stochastic process reaches the boundary set $\mathbb{B}$. 

\end{definition}
\par 
The first hitting time is also called \emph{the first exit time} of the set $\mathbb{A}$ with boundary set $\mathbb{B}$. 
It is a random variable and a stopping time of the $\sigma$ algebra family generated by the process $\bm x(t)$. We denote the mean of $t_e$ by $\overline{t}_e$. 
It is obvious that the first hitting time depends on the initial state $\bm x(0)$, the probability distribution of $\bm x(t)$ and the boundary set $\mathbb{B}$. 
\par 
In reality, the stability depends on how the system reacts to a series of small fluctuations. If we set $\mathbb{B}=\partial{\mathbb{A}}$, with the set $\mathbb{A}$ is the basin of attraction, and the initial state as the synchronous state,
the expectation of the first hitting time of  
the process $(\bm \delta(t)^\top,\bm\omega(t)^\top)^\top$ in (\ref{stochasticsystem}) can fully reflect the 
transient stability, i.e., this expectation depends on the size of the basin of 
attraction and the strength of the disturbances. This makes it a potential candidate for the metric. 
However, the distribution of the first hitting time can hardly be derived analytically or even approximated 
by the Monte-Carlo method because of the difficulty on describing 
the boundary of the basin of attraction. Due to this difficulty, we set $\mathbb{X}=\Theta$,  
\begin{align}\label{secure domain}
\Theta=\Theta_\delta\times\Theta_\omega,
\end{align}
with $\Theta_\delta$ is defined in (\ref{Theta_delta}) and
\begin{align}
  \Theta_\omega &=\{\bm\omega\in\mathbb{R}^{n}\big||\omega_{i}|<\epsilon,\forall i\in\mathcal{V}\},\label{Theta_omega}
\end{align}
where $\epsilon\in\mathbb{R}$ is a small real number corresponding to the quality of power supply according 
to the requirement of governments, i.e., the frequency fluctuation should be sufficiently small to guarantee the system stability. If the state goes out of this set, the synchronization may be lost. The set $\Theta$
is critical for monitoring the transient stability \cite{zab1}, Hence, we call it a \emph{critical set} for the
transient stability of the system in this paper. 
\par 
Let us reconsider the synchronization of the system (\ref{Nonlinear}) under the disturbances. 
With the definition of $t_e$ and $\mathbb{X}=\Theta$, the state $\bm x(t)$ of the system (\ref{stochasticsystem}) remains in the set $\Theta$ in the period $[0,t_e]$,
i.e., $\bm x(t)\in\Theta$ for any $t\in [0, t_e]$, thus the synchronization is maintained in the period $[0,t_e]$. 
Once the state exits the set $\Theta$, the synchronization may be lost.
If $t_e\rightarrow +\infty$, desynchronization will almost never happen under the disturbances. Thus, the larger 
is $t_e$, the longer is the time that the synchronization is maintained, which means the
synchronization is more stable under the disturbances. This makes the mean first hitting time $\overline{t}_e$ a 
suitable metric for the transient stability. Clearly, the mean $\overline{t}_e$ can be approximated by the Monte-Carlo method with simulations of (\ref{stochasticsystem}), thus can be used to assess the transient stability. However, to maximize
$\overline{t}_e$ in an optimization problem, one has to know the analytical expression of its probability distribution, which can hardly be derived because of the high dimension and nonlinearity of the system (\ref{stochasticsystem}). 
Thus, an alternative metric for enhancing the transient stability by an optimization framework has to be designed.

\section{CREP for transient stability} \label{Section:invariantprobability} 
\par 
In this section, we define a metric that can be minimized as an
optimization problem with the system parameters as decision variables for enhancing 
the transient stability. 
\par 
An intuitive way to increase $\overline{t}_e$ is 
to increase the probability of the state $\bm x(t)$ staying in the critical set $\Theta$. This is equivalent to increasing the 
probability of the output $\bm y(t)$ staying in the set 
\begin{align*}
\Theta_y&=\Theta_{y_\delta}\times \Theta_\omega,\\ \Theta_{y_\delta}&=\{\bm y_\delta\in\mathbb{R}^{m}\big||y_k|<\pi/2,\forall e_k=(i,j)\in\mathcal{E}\}.
\end{align*}
Due to the nonlinearity and the high dimension of the state of the system (\ref{stochasticsystem}), the probability 
distribution of the process $\bm y(t)$ can hardly be analytically obtained. Because the state $\bm x(t)$ always fluctuates around the synchronous state $\bm x^*=((\bm \delta^*)^\top,\bm 0)^\top$, the output $\bm y(t)=\bm C\bm x(t)$ fluctuates around 
the value $\bm y^*=\bm C\bm x^*$, which is seen as the expectation of the output. To investigate 
the fluctuations, the system (\ref{stochasticsystem}) is linearised around the synchronous state 
$\bm x^*=((\bm \delta^*)^\top,\bm 0)$, 
\begin{subequations}
    \label{linearization system}
    \begin{align}
    \text{d}\widehat{\bm x}(t)&=\bm A\widehat{\bm x}(t)\text{d}t+\bm B\text{d}\bm v(t),\\
    \widehat{\bm y}(t)&=\bm C\widehat{\bm x}(t),
    \end{align}
\end{subequations}
with the state variable, output vector, system matrix and input matrix
\begin{align*}
\widehat{\bm x}(t)=\begin{bmatrix}
     \widehat{\bm \delta}(t)\\
     \widehat{\bm\omega}(t)
    \end{bmatrix},~~
\widehat{\bm y}(t)=\begin{bmatrix}
    \widehat{\bm y}_{\bm\delta}(t)\\
    \widehat{\bm y}_{\bm\omega}(t)
    \end{bmatrix},~~\\
\bm A=\begin{bmatrix}
    \bm 0 & \bm I_n\\
    -\bm M^{-1}\bm L_c&-\bm M^{-1}\bm D
   \end{bmatrix},~~
\bm B =\begin{bmatrix}
    \bm 0\\
    \bm M^{-1}\widetilde{\bm B}
   \end{bmatrix},~~
\bm C&=
\begin{bmatrix}
\widetilde{\bm C}^\top&\bm 0\\
\bm 0&\bm I_n
\end{bmatrix},
\end{align*}
 where $\widehat{\bm x}$ represents the deviation of the state $\bm x(t)$ from the synchronous state $\bm x^*$, $\bm v(t)=\text{col}(v_i(t))\in\mathbb{R}^n$, $\bm M=\text{diag}(m_i),\bm D=\text{diag}(d_i),\widetilde{\bm B}=\text{diag}(b_i)$ are diagonal matrices, $\bm L_c$ is a singular Laplacian matrix whose elements satisfy
\begin{equation*}
    l_{c_{i j}}=
    \begin{cases}
       -l_{i,j}\cos(\delta_i^{*}-\delta_j^{*}), & i \neq j, \\
       \sum\limits_{k\neq i}l_{i,k}\cos(\delta_i^{*}-\delta_k^{*}), & i=j.
      \end{cases}
\end{equation*}
The matrix $\bm A$ is also called the \emph{Jacobian matrix} of the system (\ref{Nonlinear}) at the 
synchronous state $((\bm \delta^*)^\top,\bm 0)^\top$.  
Because of the Gaussian distribution of $\bm v(t)$, the process $\widehat{\bm x}$ and $\widehat{\bm y}(t)$ 
are also Gaussian such that 
\begin{align*}
\widehat{\bm x}(t)\in G\big(\bm m_{\widehat{\bm x}}(t),\bm Q_{\widehat{\bm x}}(t)\big),
~~\widehat{\bm y}(t)\in G\big(\bm m_{\widehat{\bm y}}(t),\bm Q_{\widehat{\bm y}}(t)\big).
\end{align*} 
with $\bm m_{\widehat{\bm x}}(t)\in\mathbb{R}^{2n}$ and $\bm m_{\widehat{\bm y}}(t)\in\mathbb{R}^{m+n}$,
$\bm Q_{\widehat{\bm x}}(t)\in\mathbb{R}^{2n\times 2n}$ and $\bm Q_{\widehat{\bm y}}(t)\in\mathbb{R}^{(m+n)\times (m+n)}$.
\par 
In real networks, the state always fluctuates around the synchronous state under various disturbances. 
Thus, it is reasonable to use the variance of $\widehat{\bm x}(t)$ in the invariant 
probability distribution, regardless the initial probability distribution of $\widehat{\bm x}(t)$, to 
measure the fluctuations \cite{WANG2023110884}. 
\par 
Because the system matrix $\bm A$ is singular, the invariant probability 
distribution of $\widehat{\bm x}(t)$ does not exist. However, the invariant 
probability distribution of $\widehat{\bm y}(t)$ exists \cite{WANG2023110884}, i.e., 
\begin{align*}
\lim_{t\rightarrow\infty}\bm m_{\widehat{\bm y}}(t)=\bm 0,
~~\lim_{t\rightarrow\infty}\bm Q_{\widehat{\bm y}}(t)=\bm Q_{\widehat{\bm y}}. 
\end{align*}
With this Gaussian process, we further define a new process to approximate the process $\bm y(t)$. 
\begin{definition}
Given the output $\bm y^*$ in (\ref{output}) and the Gaussian process $\widehat{\bm y}(t)$ in (\ref{linearization system}),
the stochastic process $\widetilde{\bm y}(t)$ is defined as,
\begin{align}
\widetilde{\bm y}(t)=\widehat{\bm y}(t)+\bm y^*.\label{approximateProcess}
\end{align}
\end{definition}
\par 
We denote 
$\widetilde{\bm y}(t)=(\widetilde{\bm y}_\delta^\top(t),\widetilde{\bm y}_\omega^\top(t))^\top$. The process $\widetilde{\bm y}(t)$ is Gaussian process,  
\begin{align*}
\widetilde{\bm y}(t)\in G\big(\bm m_{\widetilde{\bm y}}(t),\bm Q_{\widetilde{\bm y}}(t)\big)
\end{align*}
and has invariant probability distribution with
\begin{align*}
\lim_{t\rightarrow\infty}\bm m_{\widetilde{\bm y}}(t)=\bm y^*,
~~\lim_{t\rightarrow\infty}\bm Q_{\widetilde{\bm y}}(t)=\bm Q_{\widehat{\bm y}}. 
\end{align*} 
Clearly, $\widetilde{\bm y}(t)$ also fluctuates around $\bm y^*$. It actually approximates $\bm y(t)$ at the neighborhood 
of $\bm y^*$ because the linearisation of the system (\ref{stochasticsystem})
at the synchronous state $\bm x^*=((\bm \delta^*)^\top,\bm 0)^\top$. 
\par 
With $\bm y^*$ as the expectation solved from (\ref{syn state}) and the variance matrix $\bm Q_{\widetilde{\bm y}}$, 
the probability that the process $\widetilde{\bm y}(t)$ is in the set $\Theta$ 
at the invariant probability distribution can be calculated. 
However, this probability still can hardly be computed because an integral over a
supercube of dimension $m+n$ is needed, which involves immense computational
complexity. Thus, we focus on the marginal probability distribution of the components 
of $\widetilde{\bm y}(t)$ in the invariant probability distribution.
Denote the vector formed by the diagonal elements of the matrix $\bm Q_{\widehat{y}}$ by $\bm\sigma^2=((\bm\sigma_\delta^2)^\top,(\bm \sigma_\omega^2)^\top)^\top\in\mathbb{R}^{n+m}$ where $\bm \sigma_\delta^2=\text{col}(\sigma_{\delta_k}^2)\in\mathbb{R}^m$ and 
$\bm \sigma_\omega^2=\text{col}(\sigma_{\omega_j}^2)\in\mathbb{R}^n$ 
are the variances of the phase differences in the lines and the frequencies at the nodes. 
It is noticed that $\sigma_{\delta_k}$ and $\sigma_{\omega_j}$ 
are the standard variances of the phase difference in line $e_k$ and the frequency at node $j$ respectively. 
The logic behind enhancing the transient stability is to increase the probability 
of $\bm y(t)$ in the domain $\Theta_y$. With this logic, we define 
a critical escape probability based on the invariant marginal probability distribution of the components of $\widetilde{\bm y}(t)$ as below. 
\begin{definition}\label{metricDefinition}
Consider the stochastic process  $\widetilde{\bm y}(t)$ in (\ref{approximateProcess}). 
Denote  $\bm f=(\bm f_{\delta}^\top,\bm f_\omega^\top)^\top\in\mathbb{R}^{n+m}$ and 
$\bm f_\delta=\text{col}(f_{\delta_k})\in\mathbb{R}^m$ and $\bm f_\omega=\text{col}(f_{\omega_k})\in\mathbb{R}^m$
with  $f_{\delta_k}$ and $f_{\omega_k}$ defined as the probability of the absolute values of $\widetilde{y}_{\delta_k}$ and $\widetilde{y}_{\omega_k}$ exiting $\pi/2$ and $\epsilon$ such that 
\begin{align}
f_{\delta_k}&=1-\int_{-\pi/2}^{\pi/2}\frac{1}{\sqrt{2\pi}\sigma_{\delta_k}}e^{\frac{-(z-y^*_k)^2}{2\sigma_{\delta_k}^2}}\text{d}z,
\label{f_delta}\\
f_{\omega_k}&=1-\int_{-\epsilon}^{\epsilon}\frac{1}{\sqrt{2\pi}\sigma_{\omega_k}}e^{\frac{-z^2}{2\sigma_{\omega_k}^2}}\text{d}z,
\label{f_omega}
\end{align}
respectively. The Critical Escape Probability (CREP) assessing the probability of $\widetilde{\bm y}(t)$ 
escaping from the set $\Theta_y$ in the invariant probability distribution is defined as 
\begin{align}
\Phi=\|\bm f\|_\infty,
\end{align}
Analogously, the CREP assessing the probability of $\widetilde{\bm y}_\delta(t)$ escaping from the set $\Theta_\delta$ and 
the probability of $\widetilde{\bm y}_\omega(t)$ escaping from the set $\Theta_\omega$ are respectively defined as 
\begin{align*}
\Phi_\delta=\|\bm f_\delta\|_\infty,~~\Phi_\omega=\|\bm f_\omega\|_\infty.
\end{align*} 
\end{definition}
Because $\widetilde{\bm y}(t)$ approximates $\bm y(t)$ at the neighborhood of $\bm y^*$,
by minimizing $\Phi$, the probability of $\bm y(t)$ escaping from the critical set $\Theta$ decreases, which leads to an enhancement of the transient stability. Naturally, $\Phi$ is a metric effectively assessing the transient stability of the system (\ref{Nonlinear}) with stochastic disturbances. 
 Similarly, when \emph{the rotor angle stability} which focuses on the ability of the system to maintain the cohesiveness 
of the phase angles, and \emph{the frequency stability} which considers the severity of 
the frequency fluctuations need to be enhanced separately, $\Phi_\delta$ and $\Phi_\omega$ can 
be minimized respectively. Clearly the performance of these minimization can be evaluated by the mean first hitting time of $\bm y(t)$ to the boundary $\partial\Theta_y$, which is approximated by the Monte-Carlo method with simulations of the nonlinear 
stochastic system (\ref{stochasticsystem}).

Obviously, if $\epsilon$ in (\ref{f_omega}), which is a tolerance of the frequency fluctuations, is very large such that $\|\bm f_\delta\|_\infty>\|\bm f_\omega\|_\infty$, then
$\Phi=\Phi_\delta$. On the other hand, if $\epsilon$ is very small such that $\|\bm f_\delta\|_\infty\leq\|\bm f_\omega\|_\infty$, then $\Phi=\Phi_\omega$. 
\par 
We next present the procedure for calculating $\Phi$
and then introduce the characteristics of $\Phi$.  
\par 
The phase difference in the vector $\bm y^*$ is solved from the power flow equation (\ref{syn state}). For the calculation 
of $\bm \sigma$, we need to solve the variance matrix $\bm Q_{\widehat{y}}$ for the system (\ref{linearization system})
which is presented below.  
\par 
Because the Laplacian matrix $\bm L_c$ is symmetric, singular and semi-positive definite, we have the following lemma.
\begin{lemma}\label{appendix_theorem}
    Consider the Laplacian matrix $\bm{L_c}$ and the positive-definite diagonal matrix $\bm{M}$ in system (\ref{linearization system}). There exists an orthogonal matrix $\bm{U}\in\mathbb{R}^{n\times n}$ such that
    \begin{align}\label{decomposition}
    \bm{U}^{\bm\top}\bm{M}^{-1/2}\bm{L_c}\bm{M}^{-1/2}\bm{U}=\bm{\Lambda}_n,
    \end{align}
    where $\bm\Lambda_n=\text{diag}(\lambda_i)\in\mathbb{R}^{n\times n}$ with $0= \lambda_1< \lambda_2\cdots<\lambda_n$ being the eigenvalues of the matrix $\bm{M}^{-1/2}\bm{L_c}\bm{M}^{-1/2}$, $\bm{U}=
    \begin{bmatrix}
    \bm{u}_1&\bm{u}_2&\cdots&\bm{u}_n
    \end{bmatrix}$ with $\bm u_i\in\mathbb{R}^n$ being the eigenvector corresponding to $\lambda_i$ for $i=1,\cdots,n$. In addition, $\bm u_1=1/\sqrt{n}\bm 1_n$.
    \end{lemma}
\par
Based on Lemma \ref{appendix_theorem}, we have the following theorem \cite{WANG2023110884}.
    \begin{theorem}\label{theoremmain0}
    Consider the stochastic system (\ref{linearization system}) and
    the notations of matrices in Lemma  \ref{appendix_theorem}.
    Define matrices
    \begin{equation}
    \begin{aligned}
     &\bm{A}_e=\begin{bmatrix}
         \bm{0} & \bm{I}_n\\
        -\bm{\Lambda}_n &-\bm{U}^{\bm\top}\bm{M}^{-1}\bm{D}\bm{U}
       \end{bmatrix}\in\mathbb{R}^{2n\times 2n},\\
    &\bm{B}_e =\begin{bmatrix}
        \bm{0}\\
        \bm{U}^{\top}\bm{M}^{-\frac{1}{2}}\widetilde{\bm{B}}
       \end{bmatrix}\in \mathbb{R}^{2n\times n},\\
       &\bm{C}_e=
    \begin{bmatrix}
    \widetilde{\bm{C}}^{\bm\top} \bm{M}^{-\frac{1}{2}}\bm{U}&\bm{0}\\
    \bm{0}&\bm{M}^{-\frac{1}{2}}\bm{U}
    \end{bmatrix}\in\mathbb{R}^{(m+n)\times 2n},
        \label{blockmatrix0}
    \end{aligned}
    \end{equation}
    which can be reformulated in blocks according to
    \begin{align}\label{blockmatrix}
     \bm{A}_e=\begin{bmatrix}
             \bm{0} & \bm{A}_{12}\\
         \bm{0} &\bm{A}_2
       \end{bmatrix}, ~~
      \bm{B}_e =\begin{bmatrix}
        \bm{0}\\
        \bm{B}_2
       \end{bmatrix}, ~~
       \bm{C}_e=
       \begin{bmatrix}
       \bm{0}&\bm{C}_2
       \end{bmatrix},
    \end{align}
    where $\bm{A}_{12}\in \mathbb{R}^{1\times(2n-1)}$, $\bm{A}_2\in \mathbb{R}^{(2n-1)\times (2n-1)}$, $\bm{B}_2\in\mathbb{R}^{(2n-1)\times 2n}$, and $\bm{C}_2$ is the matrix obtained by removing the
    first column of the  matrix $\bm{C}_e$ so that
    \begin{align}\label{C2-delta-omega}
    \bm{C}_2=
    \begin{bmatrix}
    \widetilde{\bm{C}}^{\bm\top}\bm{M}^{-1/2}\widehat{\bm{U}}&\bm 0\\
    \bm{0} &\bm{M}^{-1/2}\bm{U}
    \end{bmatrix}
    \in\mathbb{R}^{(m+n)\times (2n-1)},
    \end{align}
    with $\widehat{\bm{U}}=
    \begin{bmatrix}
    \bm{u}_2&\bm{u}_3&\cdots&\bm{u}_n
    \end{bmatrix}
    \in\mathbb{R}^{n\times (n-1)}$.
    The variance matrix $\bm{Q}_{\widehat{\bm y}}$ of the output $\bm{y}$ of the system (\ref{stochasticsystem}) in the invariant probability distribution  satisfies
    \begin{align}\label{Qy}
    \bm{Q}_{\widehat{\bm y}}=\bm{C}_2\bm{Q}_{\widehat{\bm x}}\bm{C}_2^{\bm\top},
    \end{align}
    where
    $\bm{Q}_{\widehat{\bm x}}=\ds\int_{0}^{\infty}e^{\bm A_2t}\bm B_2\bm B_2^\top e^{\bm A_2^\top t}\text{d}t\in\mathbb{R}^{(2n-1)\times (2n-1)}$ that is the unique solution of the following Lyapunov equation
    \begin{align}\label{Qx}
    \bm{A}_2\bm{Q}_{\widehat{\bm x}}+\bm{Q}_{\widehat{\bm x}}\bm{A}_2^{\bm\top}+\bm{B}_2\bm{B}_2^{\top}=\bm{0}.
    \end{align}
 \end{theorem} 
 See \cite{WANG2023110884} for the proof of this theorem.
Based on Theorem \ref{theoremmain0}, we present the procedure for the calculation of $\Phi$.
 \begin{procedure}\label{procedure}
The procedure for the calculation of the metric $\Phi$, 
\begin{enumerate}[(1)]
\item Solve the power flow equation (\ref{syn state}) for the synchronous state $\bm x^*=((\bm \delta^*)^\top,\bm 0)^\top$ and output $\bm y^*=\bm C\bm x^*$;
\item Derive the stochastic process (\ref{linearization system}) by linearising the system (\ref{Nonlinear}) at the synchronous state $\bm x^*$ and model the disturbances by Gaussian noise;
\item Perform the spectral decomposition of the matrix $\bm L_c$ in Lemma \ref{appendix_theorem} and determine the 
matrices $\bm A_2, \bm B_2$ and $\bm C_2$ in Theorem \ref{theoremmain0};  
\item Solve the Lyapunov equation (\ref{Qx}) for the matrix $\bm Q_{\widehat{\bm x}}$ and calculate 
the matrix $\bm Q_{\widehat{\bm y}}$ according to (\ref{Qy}); 
\item Calculate
the vector $\bm f$ according to Definition \ref{metricDefinition} using the values of $\bm y^*$ and the diagonal element $\bm \sigma^2$ of $\bm Q_{\widehat{\bm y}}$,  
\item Calculate the norm $\|\bm f\|_\infty$. 
\end{enumerate}
\end{procedure}
\par
Clearly, $\Phi_\delta$ and $\Phi_\omega$ can be calculated at the same time with $\Phi$ by this procedure.
An important observation is that using the value of $\bm f$, the line where the system loses synchronization the most easily and the node where the frequency fluctuation is the most severe can be identified. These identifications 
allow for the discovery of weak parts in the 
network that may lead to network desynchronization.
\par 
CREP $\Phi$ has the following characteristics.
\begin{enumerate}[(i)]
\item CREP includes the influences of all the system parameters,
the inertia and the damping of the synchronous machines, the distribution of the power loads 
and generation, the line capacity, the network topology and the strength of the disturbances. 
\item CREP also reflects the size of the basin of attraction and characterizes the phenomenon that the size
of the basin shrinks if either the power flows in lines increase or if a line capacity decreases
\end{enumerate}
\par 
The first characteristic is concluded directly from the calculation of $\bm f$ and $\Phi$ in Procedure \ref{procedure}. 
Before explaining the second characteristic in a proposition, we first introduce a lemma for the bounds
of the matrix $\bm Q_{\widehat{\bm y}}$, which is needed in the proof of the proposition. 
For matrices $\bm A, \bm B\in\mathbb{R}^{n\times n}$, we say that $\bm A\preceq \bm B$ if 
the matrix $\bm A-\bm B$ is semi-negative-definite. 
To emphasize the variance matrix of the frequencies and the 
phase differences, we write the matrix $\bm Q_{\widehat{\bm y}}$ in the following form,
\begin{align}\label{blockformofQy}
\bm{Q}_{\widehat{\bm y}}=\begin{bmatrix}
             \bm{Q}_{\widehat{\delta}} & \bm{Q}_{\widehat{\delta}\widehat{\omega}}^\top\\
         \bm{Q}_{\widehat{\delta}\widehat{\omega}} &\bm{Q}_{\widehat{\omega}}
       \end{bmatrix}\in\mathbb{R}^{(m+n)\times (m+n)}.  
\end{align}
\begin{lemma}\label{Lemma_bounds}
Consider the stochastic process (\ref{linearization system}). Define $\overline{\eta}=\max\{\eta_i,i=1,\cdots,n\}$ and 
$\underline{\eta}=\min\{\eta_i,i=1,\cdots,n\}$ with $\eta_i=b_i^2/d_i$. 
The variance matrix $\bm Q_{\widehat{\delta}}$ satisfies
\begin{align}\label{boundsofQ}
\frac{1}{2}\underline{\eta}\bm S\preceq \bm Q_{\widehat{\delta}}\preceq
\frac{1}{2}\overline{\eta}\bm S,~~\bm S=\bm {\widetilde{C}^\top M^{-1/2}\widehat{U}\Lambda_{n-1}^{-1}\widehat{U}^\top M^{-1/2}\widetilde{C}}
\end{align}
where $\Lambda_{n-1}=\text{diag}(\lambda_i,i=2,\cdots,n)\in\mathbb{R}^{(n-1)\times(n-1)}$. 
\end{lemma}
\emph{Proof:} 
Define matrices 
$\overline{\bm \beta}=(\overline{\eta}\bm D)^{1/2}$ and 
$\underline{\bm \beta}=(\underline{\eta}\bm D)^{1/2}$
and 
\begin{align*}
\bm Q_{\overline{\beta}}=
\int_{0}^{\infty}e^{\bm A_2t}\overline{\bm B}_2\overline{\bm B}_2^\top e^{\bm A_2^\top t}\text{d}t,~
\bm Q_{\underline{\beta}}=
\int_{0}^{\infty}e^{\bm A_2t}\underline{\bm B}_2\underline{\bm B}_2^\top e^{\bm A_2^\top t}\text{d}t
\end{align*}
with $\overline{\bm B}_2, \underline{\bm B}_2\in\mathbb{R}^{(2n-1)\times n}$ such that 
\begin{align*}
\overline{\bm B}_2=
\begin{bmatrix}
\bm 0\\
\bm U^{{\bm\top}}\bm M^{-1/2}\overline{\bm \beta}
\end{bmatrix},~~
\underline{\bm B}_2=
\begin{bmatrix}
\bm 0\\
\bm U^{{\bm\top}}\bm M^{-1/2}\underline{\bm \beta}
\end{bmatrix}.
\end{align*}
From the definition of $\overline{\bm \beta}$ and $\underline{\bm \beta}$ and $\underline{\eta}d_i\leq b_i^2=\eta_id_i\leq \overline{\eta}d_i$ for all the nodes,
it yields
\begin{align*}
\underline{\eta}\text{diag}(d_i)=\underline{\bm \beta}\underline{\bm \beta}^{\bm\top}\preceq \widetilde{\bm B}\widetilde{\bm B}^{\bm\top}=\text{diag}(b_i^2) \preceq \overline{\bm \beta}\overline{\bm \beta}^{\bm\top}=\overline{\eta}\text{diag}(d_i).
\end{align*}
which leads to 
\begin{align*}
\underline{\bm B}_2\underline{\bm B}_2^{\bm\top}\preceq \bm B_2\bm B_2^{{\bm\top}}\preceq \overline{\bm B}_2\overline{\bm B}_2^{\bm\top}
\end{align*}
By Theorem \ref{theoremmain0}, we further obtain
\begin{align}\label{sec6-2}
\bm Q_{\underline{\beta}}\preceq \bm Q_{\widehat{x}}\preceq \bm Q_{\overline{\beta}}.
\end{align}
Following \cite[Lemma 4.2]{WANG2023110884}, we derive the explicit formula 
of $\bm Q_{\underline{\beta}}$ and $\bm Q_{\overline{\beta}}$
by solving the corresponding Lyapunov equations respectively,
\[
\bm Q_{\underline{\beta}}=
\begin{bmatrix}
\frac{1}{2}\underline{\eta}\bm \Lambda_{n-1}^{-1}&\bm 0\\
\bm 0&\frac{1}{2}\underline{\eta}\bm I
\end{bmatrix}
, ~~
\bm Q_{\overline{\beta}}=
\begin{bmatrix}
\frac{1}{2}\overline{\eta}\bm \Lambda_{n-1}^{-1}&\bm 0\\
\bm 0&\frac{1}{2}\overline{\eta}\bm I
\end{bmatrix}.
\]
With these explicit formulas and (\ref{Qy}) and the form in (\ref{blockformofQy}), we
obtain (\ref{boundsofQ}). \hfill $\square$

\par 
Based on this Lemma, we have the following theorem. 
\begin{theorem}
Consider CREP $\Phi$ in Definition \ref{metricDefinition}. It holds that
\par
(1) each element of the vector $\bm f$ and the value $\Phi$ are in the interval $[0,1]$; 
\par
(2) if the second smallest eigenvalue of the matrix $\bm L_c$ at the
synchronous state decreases to zero, then the metric $\Phi$ increases to one. 
\end{theorem}
\emph{Proof:}
(1) At a synchronous state, when the strength of the
disturbances vary from zero to infinity, the variance $\bm \sigma_\omega^2$ of 
the frequencies at the nodes and $\bm \sigma_\delta^2$ in the lines  vary from zero to infinity. 
It follows from Definition \ref{metricDefinition} for $f_{\delta_k}$ and $f_{\omega_k}$,
the values of 
$\Phi$ lies in the interval $[0,1]$. 
\par
(2)
By the definition in (\ref{f_delta}), $f_{\delta_k}$ decreases to one as the variance $\sigma_{\delta_k}$
increases to infinity. With the bounds of the matrix $\bm Q_{\widehat{\delta}}$ in Lemma \ref{Lemma_bounds}, we only need to prove that as the second smallest eigenvalue decreases to zero, there is at least one diagonal element of the matrix $\bm S$ that increases to infinity. The incidence matrix of the graph is written into 
$\bm C=
\begin{bmatrix}
\bm c_1&\bm c_2 &\cdots &\bm c_m
\end{bmatrix}$, where $\bm c_k$ describes the indices of the two nodes connected by line $e_k$.  Without loss of generality, assume the line $e_k$ connects nodes $i$ and $j$ and the direction of this line is from node $i$ to $j$. Then, the $i$-th and $j$-th element of the vector $\bm c_k$ are $c_{ik} = $1 and $c_{jk}=-1$, respectively and the other elements all equal to zero. From the definition of the matrix $\bm S$ in Lemma \ref{Lemma_bounds}, we obtain the diagonal elements of $\bm S$,
\begin{equation*}
    s_{kk}=\sum_{q=1}^{n-1}\lambda_{q+1}^{-1}(m_{i}^{-1/2}u_{i,q+1}-m_{j}^{-1/2}u_{j,q+1})^2,\,\,k=1,2,\cdots,m,
\end{equation*}
where $u_{i,q+1}$ and $u_{j,q+1}$ are the $i$-th and $j$-th element of the vector $\bm u_{q+1}$. Here $\bm u_{q+1}$ is the $(q+1)$-th column of the matrix
$\bm U$ defined in Lemma \ref{appendix_theorem}. Because $\bm u_2$ is a column of the orthogonal matrix $\bm U$, there exists $i,j$ with $i\neq j$ such that $m_{i}^{-1/2}u_{i,2}\neq m_{j}^{-1/2}u_{j,2}$, thus $s_{kk}$ increases to infinity as the second smallest eigenvalue $\lambda_2$ decreases to zero. \hfill $\square$
\par
This theorem indicates that CREP $\Phi$ fully reflects
the size of the basin of attraction of a stable synchronous state. 
In fact, it is known that as the power loads increases or the 
line capacities decreases, the synchronous state $((\bm \delta^*)^\top,\bm 0)^\top$ moves to the boundary $\partial \Theta$ and both the second smallest eigenvalue $\lambda_2$
of $\bm L_c$ and the size of the basin of attraction decrease. For the system (\ref{Nonlinear}), the number of 
eigenvalues of its system matrix $\bm A$ with positive 
real part equals to the number of negative eigenvalues
of $\bm L_c$ \cite{zab1}. Thus, when the secondary smallest eigenvalue
of $\bm L_c$ decreases to zero, the stable synchronous state gradually disappears, which 
means the basin of attraction disappears. Clearly, this is captured by CREP $\Phi$ which
increases to one in this case. 
This theorem also demonstrates that 
CREP fully reflects the phenomena that if
the synchronous state is close to the boundary, 
a very small disturbance may lead to desynchronization. 
\par
To illustrate the procedure for calcluating 
CREP and its characteristics, we apply it to the \emph{Single Machine Infinite Bus} (SMIB) model with Gaussian disturbances, 
\begin{example}
Consider the SMIB model with Gaussian disturbances, 
\begin{align*}
\dot{\delta}(t)&=\omega(t),\\
M\dot{\omega}(t)&=P-D\omega(t)-K\sin{\delta}(t)+bw(t).
\end{align*} 
\end{example}
Assume $P\leq K$, obviously, an equilibrium point is $(\delta^*,\omega^*)=(\arcsin{P/K},0)$. 
Linearising the system at the equilibrium point,
we obtain a Gaussian stochastic process
with system matrix and input matrix 
\begin{align*}
\bm A=\begin{bmatrix}
 0&1\\
-M^{-1}l_c&-M^{-1}D
\end{bmatrix},~
\bm B=\begin{bmatrix}
0\\
M^{-1}b
\end{bmatrix}
\end{align*} 
where $l_c=K\cos{\delta^*}=\sqrt{K^2-P^2}$. By solving the following 
Lypunov equation,
\begin{align*}
\bm A\bm Q+\bm Q\bm A^\top+\bm B\bm B^\top=\bm 0, 
\end{align*} 
we obtain the variance matrix in the invariant probability distribution of the stochastic process,
\begin{align*}
\bm Q=\begin{bmatrix}
\frac{b^2}{2D\sqrt{K^2-P^2}}&0\\
0&\frac{b^2}{2MD}
\end{bmatrix}.
\end{align*}
With this variance matrix and the equilibrium point as the expectation, the critical probability $f_\delta$ and 
and  $f_\omega$ are calculated according to Definition \ref{metricDefinition}. 
Clearly, $f_\omega$ depends on the inertia and the
damping of the synchronous machines and the strength of the disturbance while independent 
of the line capacity $K$ and the load $P$. However, the dependence of $f_\delta$ on the system parameters is relatively complex. 
\begin{figure}
\centering
\includegraphics[scale=1.02]{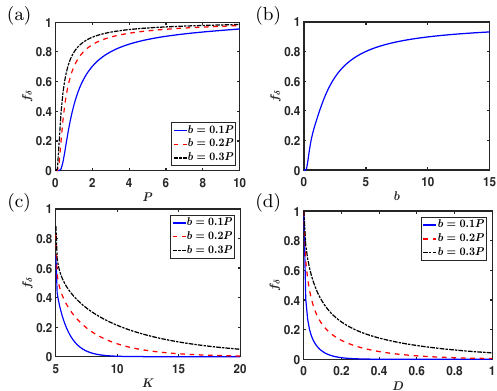}
\caption{The dependence of $f_\delta$ on the system parameters in SMIB model. }\label{fig1}
\end{figure}
\par 
It is clear that $f_\delta$ is independent of the inertia. 
Fig.  \ref{fig1} shows the trend of $f_\delta$ as the system parameters changes. 
In particular,  as the load $P$ increases to the line capacity, the basin of attraction 
of the equilibrium $(\delta^*, 0)$ gradually disappears. This is fully confirmed
by $f_\delta$, which increases to one.  In Section \ref{Section:Case study},
the  Braess' parodox will be revisited with the proposed metric, where new findings will be presented.

\section{The optimization framework}\label{Section:Optimization Framework}

The proposed metric CREP actually quantifies the risk that the state escape 
from the critical set $\Theta$. By minimizing this metric with the system parameters
as the decision variables, this risk can be decreased, thus increasing the transient stability.  
For this minimization, we introduce an
optimization framework with the choice of the decision variables as 
the line capacities, the power generations, the inertia coefficients and the damping coefficients,
\begin{align}
    &\min_{\bm \theta}~~ \Phi=\|\bm f\|_\infty,\label{optimize}\\
    \nonumber
    \text{s.t.} ~~&\text{(\ref{syn state}),~(\ref{output}),~(\ref{f_delta}),(\ref{f_omega}),(\ref{decomposition}), (\ref{Qy}), (\ref{Qx})},
    \\
    &|\delta_i^*-\delta_j^*|\leq \pi/2, (i,j)\in\mathcal{E},\label{ConstaintPhase}
     \\
    &\bm g(\bm \theta)\leq 0, \label{decisionconstraint}
\end{align}
where $\bm \theta$ denotes the selected decision variable, $\bm g(\bm \theta)\leq 0$ denotes the constraint on the selected decision variables.
 If only the rotor angle stability is considered, the objective function is replaced by
$\Phi_\delta$. Similarly, if suppressing 
the frequency fluctuations is the main purpose, the objective function is replaced by 
$\Phi_\omega$. 
We remark that the constraints (\ref{ConstaintPhase}) cannot be 
neglected because there may be many equilibrium points for the system (\ref{Nonlinear}) that 
are not in the set $\Theta$ \cite{Xi2016} and unexpected unstable synchronous state may be obtained if these 
constraints are neglected.  
\par 
If the power generation are selected as the decision variables, which is usually optimized in
the tertiary frequency control, the constraints (\ref{decisionconstraint}) are 
replaced by 
\begin{subequations}\label{OptimalGeneration}
\begin{align}
 &0=\text{P}_t-\sum\limits_{i\in\mathcal{V}_g} P_i,\\ 
    &\underline{P}_i\leq P_i\leq \overline{P}_i, i\in\mathcal{V}_g, \label{BoundConstraints1}
\end{align}
\end{subequations}
where $P_t$ is the total power load, $\mathcal{V}_g\subset\mathcal{V}$ denotes 
the set of power generations, $\underline{P}_i$ and $\overline{P}_i$ are the lower bound and upper bound of the power generation
respectively. Note that the power load at a node may also be selected as a decision variable, the form of 
the constraints are the same as (\ref{OptimalGeneration}). 
\par
If the inertia of the synchronous machines are the decision variables such that $\bm \theta=\text{col}(m_{i})\in\mathbb{R}^{n}$, the constraints (\ref{decisionconstraint})
are replaced by 
\begin{subequations}\label{OptimalInertia}
\begin{align}
    &0=\text{M}_t-\sum\limits_{i=1}^n {m_{i}},\\
    &\underline{m}_i\leq m_i\leq \overline{m}_i, i=1,\cdots,n,\label{BoundConstraints2}
\end{align}
\end{subequations}
where $\text{M}_t$ denotes the total amount of inertia and $\underline{m}_i$ and $\overline{m}_i$ are the lower bounds and upper bounds of inertia coefficients respectively.
\par 
Similarly, if the damping coefficients are selected as decision variables
such that $\bm \theta=\text{col}(m_{i})\in\mathbb{R}^{n}$, the constraints (\ref{decisionconstraint}) 
are then replaced by 
\begin{subequations}\label{OptimalDamping}
\begin{align}
    &0=\text{D}_t-\sum\limits_{i=1}^n {d_{i}},\\
    &\underline{d}_i\leq d_i\leq \overline{d}_i, i=1,\cdots,n,\label{BoundConstraints3}
\end{align}
\end{subequations}
where $\text{D}_t$ denotes the total amount of damping and $\underline{d}_i$ and $\overline{d}_i$ are the lower bound and upper bound of the damping coefficient at node $i$ respectively.
\par 
If the line capacities of the lines are selected as decision variables, 
i.e., $\bm \theta=\text{col}(l_{i,j})\in\mathbb{R}^m$ with $(i,j)\in\mathcal{E}$,
the constraints 
(\ref{decisionconstraint}) are replaced by 
\begin{subequations}\label{OptimalCapacity}
\begin{align}
&0=\text{L}_t-\sum_{(i,j)\in\mathcal{E}}{l_{i,j}},\\
&\underline{l}_{i,j}\leq l_{i,j}\leq \overline{l}_{i,j}, (i,j)\in\mathcal{E}, \label{BoundConstraints4}
\end{align}
\end{subequations}
where $\text{L}_t$ is the total available line capacities and $\underline{l}_{i,j}$ and 
$\overline{l}_{i,j}$ are the lower bound and the upper bound of the capacity of line $(i,j)\in\mathcal{E}$ respectively. 

\par 
When the objective function in (\ref{optimize}) is replaced by $\Phi_\omega=\|\bm f_\omega\|_\infty$ , 
the frequency fluctuations are minimized by tuning the system parameters. 
For this minimization problem, we have the following proposition.
\begin{proposition}\label{metricEquivalent}
Consider the metric $\Phi_\omega=\|\bm f_\omega\|_\infty$ with $\bm f_\omega$ defined 
in Definition \ref{metricDefinition}. Minimizing $\|\bm f_\omega\|_\infty$
is equivalent to minimizing $\|\bm \sigma_\omega^2\|_\infty$. 
\end{proposition}
\emph{Proof:}
It holds
\begin{align*}
    \frac{\text{d}f_{\omega_k}}{\text{d}\sigma_{\omega_k}}&=-\int_{-\epsilon}^{\epsilon}\frac{1}{\sqrt{2\pi}\sigma_{\omega_k}^2}e^{-\frac{z^2}{2\sigma_{\omega_k}^2}}\left(
        \frac{z^2}{\sigma_{\omega_k}^2}-1
    \right)\text{d}z\\
    &=\frac{1}{\sqrt{2\pi}\sigma_{\omega_k}^{2}}ze^{-\frac{z^2}{2\sigma_{\omega_k}^2}}\bigg | _{z=-\epsilon}^{z=\epsilon}
    =\frac{2\epsilon}{\sqrt{2\pi}\sigma_{\omega_k}^2}e^{-\frac{\epsilon^2}{2\sigma_{\omega_k}^2}}>0. 
\end{align*}
The value $f_{\omega_i}$ is monotonically increasing with respect to the standard variance $\sigma_{\omega_i}$. 
Thus, minimizing $\|\bm f_\omega\|_\infty$
is equivalent to minimizing $\|\bm \sigma_\omega^2\|_\infty$.  \hfill $\square$

With this proposition, the objective function can be further 
replaced by $\|\bm \sigma_\omega^2\|_\infty$. We remark
that this is different from minimizing the $\mathcal{H}_2$ norm as in the optimization framework 
(\ref{optimalNorm}) where the objective function is the sum of the frequency variances 
at all the nodes. 
\par 
Note that in the constraints (\ref{OptimalGeneration}), (\ref{OptimalInertia}), (\ref{OptimalDamping})
and (\ref{OptimalCapacity}), 
the upper bound may equal to the lower bound, in which case the corresponding decision variables
become constants. The constraints (\ref{ConstaintPhase}) restrict the synchronous state in the domain (\ref{secure domain}). 
Because the synchronous state may not exist, there may be no 
solutions for the optimization problem in that case.

\section{Case study}\label{Section:Case study}
\begin{figure}[ht]
    \centering
	\includegraphics[width=0.32\textwidth]{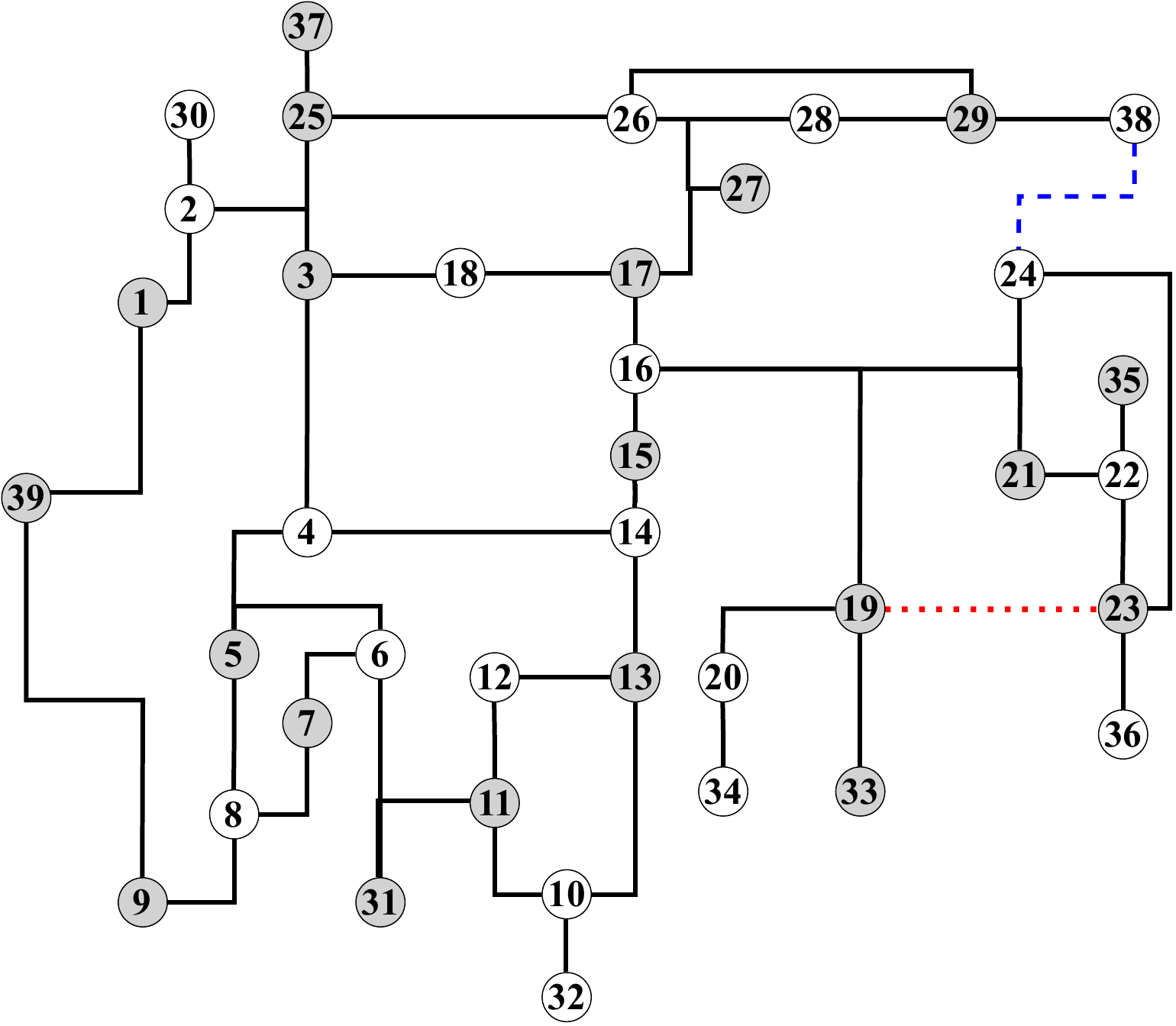}
    \caption{A network for case study. The grey nodes represent power loads and the blank nodes represent power generations.}\label{figure1}
\end{figure}

In this section, we evaluate the performance of the proposed metric on 
assessing the transient stability and the optimization framework on enhancing the transient stability of a system  with network topology as shown in Fig. \ref{figure1}.
In this model, all the buses are assumed to be synchronous machines.
There are 39 nodes and 
46 lines. The nodes with even numbers are connected to power generators and the other nodes are connected 
to power loads, which are denoted by blank nodes and grey nodes respectively in Fig. \ref{figure1}.
\par 
Because the solution of the optimization problem with objective (\ref{optimize}) 
is sensitive to the parameter $\epsilon$, i.e., if it is too large, 
the phase difference will often first hit the boundary, while if it is too small,
the frequency component will often first hit the boundary. Due to difficulty in the configuration, we first study the metric $\Phi$ regardless
the frequency fluctuations, which lead to $\Phi=\Phi_\delta$, and then study it without boundary trigger of the phase differences, which leads to $\Phi=\Phi_\omega$. For the former case,
we only need to evaluate the performance of the metric $\Phi_\delta=\|\bm f_\delta\|_\infty$ and the corresponding 
optimization framework. 
For the latter one, we evaluate $\Phi_\omega=\|\bm f_\omega\|_\infty$ and its corresponding optimization framework. 
In particular, we revisit 
the Braess' paradox, in which according to earlier studies using different metrics the stability may be decreased when a
 new line is added or the line capacity of an existing line is increased. 
\par 
The phase cohesiveness measured either by $\|\bm y^*_\delta\|_\infty$ or the order parameter at the synchronous state and the $\mathcal{H}_2$ norm of the system with 
stochastic input may be considered as metrics for optimal network design\cite{FAZLYAB2017181,Skardal2014}, for which 
the corresponding optimization frameworks are introduced in the Appendix. Here, we 
compare the performance of these optimization frameworks to 
that of the proposed framework in this paper.  The corresponding optimization problems are solved by 
the Genetic Algorithm method using Matlab. The bound constraints (\ref{BoundConstraints1}-\ref{BoundConstraints4}) of the decision variables
are not considered.
\par 
To show the results intuitively, 
the mean first hitting time $\overline{t}_e$ of $\bm y(t)$ to the boundary $\partial \Theta_y$ is used to indicate
the enhancement of the transient stability in these evaluations, which is calculated statistically by the Monte-Carlo method for the nonlinear system (\ref{stochasticsystem}). 
The Euler-Maruyama method 
is applied to the system (\ref{stochasticsystem}) with the initial condition $(\bm \delta(0), \bm \omega(0))=(\bm \delta^*,\bm 0)$ and simulation time $T=10^5$. The total number of samples for calculating 
the mean first hitting time is $N=10^5$ and the time step for 
the simulation is $\Delta t=10^{-3}$.

In subsection \ref{optimize-phase}, we investigate the dependence of $\Phi_\delta$ on the 
system parameters and the relationship between $\Phi_\delta$ and  $\overline{t}_e$. 
The performance of minimizing $\Phi_\delta$ and the revisit of Braess' paradox are also 
described in this subsection. In subsection \ref{optimize-omega}, we introduce the 
dependence of $\Phi_\omega$ on the system parameters and the performance 
of minimizing $\Phi_\omega$.

\subsection{The metric $\Phi_\delta$}\label{optimize-phase}
To understand
the dependence of the metric $\Phi_\delta$ on the system parameters and the performance of proposed optimization framework,  we focus the systems with the following 5 configurations of system parameters respectively, where 
the decision variables may be the power generation, the line capacities of all the lines,
the inertia and the damping coefficients of the synchronous machines at all the nodes separately. 
\begin{enumerate}[(1)]
\item The parameters selected as decision variables are set identically.
For example, when the power generation is selected as decision variables, we 
set all the power generation identically with total power supply $\text{P}_t$ in these systems, i.e., $P_i=P_t/N_g$ where 
$N_g$ is the total number of generators and also is the dimension of the decision variables.  
We mention the models with this parameter configuration as \emph{initial models} for simplicity;
\item  The parameters selected as decision variables are set to the solution of the optimization problems minimizing $\Phi_\delta$. For example, when the power generation are selected as decision variables, the values
of the power generation is set to the solution of the optimization problems with objective $\Phi_\delta$. 
\item  The parameters selected as decision variables are set to the solution of the  optimization problems minimizing  $\text{tr}(\bm Q_{\widehat{\delta}})$
which actually is the $\mathcal{H}_2$ norm of the system (\ref{linearization system}); See (\ref{optimalNorm}) for the 
corresponding optimization problems; 
\item The parameters selected as decision variables are set to the solution of the optimization problems minimizing $\|\bm y^*_\delta\|_{\infty}$, which measures 
the cohesiveness of the phases. See (\ref{optimalinfinity}) for the corresponding optimization problems. 
\item  The parameters selected as decision variables are set to the solution of the optimization problems maximizing the order parameter, which is  often used to study the level of the synchronization of complex network systems. See (\ref{optimalOrder}) for the corresponding optimization problems. 
\end{enumerate}
 We set $b_i=0.5\sqrt{i+1}$. 
If a parameter is not selected as a decision variable, it is set as in Table \ref{Parameters}.  
For example, if the power generation are selected as 
decision variables, then we set the other variables as shown in Table \ref{Parameters},
i.e., $l_{i,j}=20$, $M_i=0.08i$, $d_i=0.2\times (42-i)$.
Specially, when studying the dependence of the metric $\Phi_\delta$
and the mean first hitting time $\overline{t}_e$ on the power generation and loads, we set the power loads identically with total amount 
of power $\text{P}_t$.  In the Monte-Carlo methods 
for the simulations of the system (\ref{stochasticsystem}), the first hitting 
time $t_e$ is recorded when there are lines in which the phase differences 
exit the set $\Theta_\delta$ regardless of the deviations of the frequencies.
The initial models are used for comparing
the performance of the 4 optimization frameworks.

\subsubsection{The dependence of $\Phi_\delta$ on the system parameters}
 
\begin{figure}
    \centering 
    \includegraphics[scale=1.05]{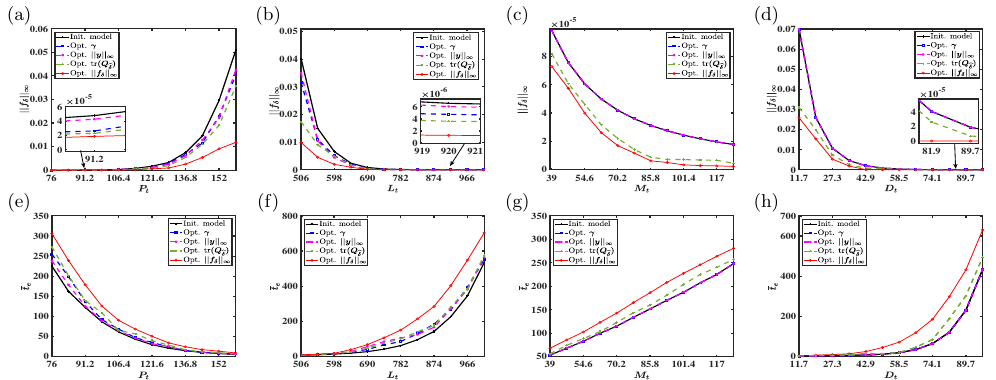}
    \caption{The dependence of $\overline{t}_e$ and $\|\bm f_\delta\|_\infty$ 
    on the parameters $\text{P}_t$, $\text{L}_t$, $\text{M}_t$ and $\text{D}_t$ in the models with the 5 configurations of paremeters.   }\label{fig.phaseDifference}
\end{figure}

The dependence of $\overline{t}_e$ and $\Phi_\delta$ on the parameters $\text{P}_t$, $\text{L}_t$ $\text{M}_t$ and $\text{D}_t$ are shown in Fig. \ref{fig.phaseDifference}. The findings from these figures are summarized below. 
\par 
First, by comparing the trends of $\Phi_\delta=\|\bm f_\delta\|_\infty$ as these parameters change in Fig. \ref{fig.phaseDifference}(a-d)
with those of $\overline{t}_e$ in Fig. \ref{fig.phaseDifference}(e-h), it can be observed that 
when the metric $\|\bm f_\delta\|_\infty$ increases, $\overline{t}_e$ 
decreases. 
This demonstrates that CREP fully
reflects the trends of the mean first hitting time, and shows the ability of  effectively assessing the transient stability in terms of the mean first hitting time. 
\par 
Second, it is found from Fig. \ref{fig.phaseDifference}(a,b,d)
that $\Phi_\delta=\|\bm f_\delta\|_\infty$ decreases as 
$\text{L}_t$ and $\text{D}_t$ increase respectively while increases
as $\text{P}_t$ increases in all the 5 models. This is practical. In particular, it is shown in Fig. \ref{fig.phaseDifference}(c) 
that when the inertia increases, $\|\bm f_\delta\|_\infty$ decreases significantly. This demonstrates
that increasing the inertia is also beneficial to the rotor angle stability,
which is consistent with the findings from the explicit formula 
of the variance matrix of the phase differences for star networks in \cite{WuXian2}. This is because 
a large inertia accelerates the propagation of the disturbances from a node
to the other nodes. It is remarked that with the assumption of 
uniform disturbance-damping ratio, i.e., $b_i^2/d_i=\eta$ for all the 
nodes, the variance of the phase differences is independent 
of the inertia \cite{WANG2023110884}. 
\par 
We remark that the objectives $\|\bm y^*_\delta\|_\infty$ in (\ref{optimalinfinity}) and 
the order parameter (\ref{optimalOrder}) are independent of the inertia 
and damping coefficients of synchronous machines. Thus, when 
the inertia and the damping coefficients are selected as decision variables, 
the values of $\overline{t}_e$ will not be changed. This is shown 
in Fig. \ref{fig.phaseDifference}(g,h) where the curves of the mean 
first hitting time in the initial model and the ones with system parameters setting to the solutions 
of maximizing the order parameter $\gamma$ and minimizing $\|\bm y^*_\delta\|_\infty$ coincide.


\subsubsection{Performance of minimizing $\Phi_\delta$ }

We compare the performances of the optimization frameworks, i.e.,
the ability on increasing the mean first hitting time $\overline{t}_e$.
In Fig. \ref{fig.phaseDifference}(e-h), it is clearly shown
that $\overline{t}_e$ with system parameters optimized by the proposed optimization framework, that are denoted by red dotted lines, is much 
larger than all the others. 
This demonstrates that minimizing $\Phi_\delta$ is more effective on 
increasing the transient stability than optimizing all the other metrics.
\emph{This also confirms that for enhancing the transient stability, it is insufficient to suppress the fluctuations only.} Obviously, when the strength of the disturbance decreases to zero which 
leads $\bm \sigma_\delta^2$ to zero, 
the effectiveness gradually reduces to that of the optimization framework (\ref{optimalinfinity}).

\subsubsection{Revisit of the Braess' paradox with $\Phi_\delta$}

If a new line is added or the capacity of a line increases, 
its influences can be evaluated from the changes of the linear stability measured
by the absolute values of the real parts of the non-zero eigenvalues of 
the system matrix $\bm A$ in (\ref{linearization system}) or the order parameter $\gamma$ defined in (\ref{orderparameter}).
We denote the smallest absolute value of the real parts of the nonzero eigenvalues by 
 $\min\{|\text{Re}(\mu_i)|\}$ where $\mu_i$ denotes the non-zero eigenvalues of $\bm A$. 
A Braess' paradox occurs if $\min\{|\text{Re}(\mu_i)|\}$ or the order parameter $\gamma$ decrease 
when a new line is added or the capacity of a line increases. 
Here, on the network shown in Fig. \ref{fig1}, we study the performance of $\Phi_\delta$  on 
identifying a Braess' paradox and compare it with those of the linear stability and the order parameter. 
We set $b_i=0.09i$ and set the other parameters as in Table \ref{Parameters}. 
We show in Table \ref{paradox} the values of $\overline{t}_e$, $\|\bm f_\delta\|_\infty$, $\min\{|\text{Re}(\mu_i)|\}$ and $\gamma$, where the confidence intervals of $\overline{t}_e$ with confidence level $95\%$ are $[\overline{t}_e-t_c,\overline{t}_e+t_c]$ with $t_c\leq 2\text{s}$ in all the 4 cases. 
 
\par 
Let us first focus on the changes of these metrics after a new line is added, i.e., either 
line $(19,23)$ in case 2 or $(24,38)$ in case 3. It is shown 
in Table \ref{paradox} that after adding line $(19,23)$, $\min\{|\text{Re}(\mu_i)|\}$ and $\gamma$
increase from $0.2833$ to $0.3179$ and $0.9663$ to $0.9666$ respectively, which indicate
the stability is increased. While, $\|\bm f_\delta\|_\infty$ increases
from  $4.383\times 10^{-6}$ to $1.419\times 10^{-5}$ and $\overline{t}_e$ decreases from $195.14$s to $148.94$s, both indicate the stability is  
decreased. Clearly, this conflicts with the result by $\min\{|\text{Re}(\mu_i)|\}$ and $\gamma$. 
Conversely, in case of adding line $(24,38)$, a Braess' paradox is identified 
with respect to $\min\{|\text{Re}(\mu_i)|\}$ and $\gamma$, which decrease from $2.833$ to $2.832$ 
and from $0.9663$ to $0.9652$ respectively. In contrast, with the metric $\|\bm f_\delta\|_\infty$, which 
decreases from $4.383\times 10^{-6}$ to $3.393\times 10^{-6}$, and the metric $\overline{t}_e$, which 
increases from $195.14$s to $241.51$s, it is identified that the new added line increases the stability. 
\par

We next study the changes of these metrics after increasing the line capacity of $(22,35)$
by comparing the results of case 1 and case 4 in Table 
\ref{paradox}. 
It is seen that after increasing the line capacity, $\min\{|\text{Re}(\mu_i)|\}$ and 
the order parameter increase from $0.2833$ to $0.3103$ and from $0.9663$ to $0.9666$ respectively,
both indicate that increasing the line capacity of $(22,35)$ is beneficial 
to the stability. However, the metric $\|\bm f_\delta\|_\infty$ increases 
from $4.383\times 10^{-6}$ to $4.967\times 10^{-6}$ and the mean first hitting time decreases from $195.14$s to $188.11$s, which indicate that stability decreases, thus a Braess' paradox occurs. 
\par 
In words, \emph{whether a Braess' paradox occurs depends on the metric used 
for the stability.} The proposed metric that involves 
the roles of all the system parameters and the strength of 
disturbances provides a more practical tool to identify a Braess' paradox.

\subsection{The metric $\Phi_\omega$ }\label{optimize-omega}
In this subsection, 
we study the dependence of the metric $\Phi_\omega=\|\bm f_\omega\|_\infty$ on
the system parameters and the performance of the optimization framework. 
We focus on the mean first hitting time $\overline{t}_e$ when the frequencies exit 
the range $\Theta_\omega$ and $\|\bm f_\omega\|_\infty$
in the systems with the following 3 configurations of system parameters. 
\begin{enumerate}[(1)]
\item The parameters selected as decision variables are set identically.  The model 
with this parameter configuration are also called \emph{initial models} and used for comparison as in the previous subsection. 
\item The parameters selected as decision variables are set to the 
solution of the optimization problems minimizing tr($\bm Q_{\widehat{\omega}}$).
\item The parameters selected as decision variables are set to the 
solution of the optimization problems minimizing $\|\bm \sigma^2_\omega\|_\infty$.
For the optimization problems, see (\ref{optimalNorm}) with the objective replaced by $\|\bm \sigma^2_\omega\|_\infty$.
\end{enumerate}
\par 
 We set $b_i=5 i\times 10^{-4}$, which is much smaller than 
the setting in Subsection \ref{optimize-phase}.
The parameters that are not selected as decision variables are set to the values in Table \ref{Parameters}. 
Note that minimizing $\|\bm \sigma^2_\omega\|_\infty$ is equivalent to minimizing $\|\bm f_\omega\|_\infty$
based on Proposition \ref{metricEquivalent}. 
As in the previous subsection, when studying the impact of the 
power generation and loads on the metric $\|\bm f_\omega\|_\infty$, we select 
all the power generation as decision variables and set the power load identically 
with the total amount $\text{P}_t$.  We set $\epsilon=0.02$ to calculate $\|\bm f_\omega\|_\infty$ and $\overline{t}_e$
in the simulations of (\ref{stochasticsystem}).  In the Monte-Carlo methods 
for the simulations of the system (\ref{stochasticsystem}), the first hitting 
time is recorded when there are nodes at which the frequencies 
exit the range $\Theta_\omega$ regardless of the fluctuations of the phase differences. 
Note that in all the simulations, because the strengths of the disturbances
are much smaller than those in Subsection \ref{optimize-phase}, the phase differences
in all the lines remain in the range $\Theta_\delta$ in the simulations.  
In other words, the frequencies always hit the boundary of $\Theta_\omega$ first. 
The simulation results are shown in Fig. \ref{fig.frequency}.

\subsubsection{The dependence of $\Phi_\omega$ on the system parameters}
\par 
It is observed from Fig. \ref{fig.frequency}
that $\|\bm f_\omega\|_\infty$ increases as $\text{P}_t$
increases and decreases as $\text{L}_t$ increases.
This is because either increasing $\text{P}_t$ or decreasing
$\text{L}_t$ will decrease the weight $l_{i,j}\cos(\delta_i^*-\delta_j^*)$
for all $(i,j)\in\mathcal{E}$, which decelerates the propagation 
of disturbances from a node to the others. 
Note that accelerating the propagation of the disturbances
in a network with heterogeneous strength of disturbances is beneficial 
to decrease $\|\bm \sigma_\omega^2\|_\infty$ which further decreases
 $\|\bm f_\omega\|_\infty$. This is consistent with the theoretical 
 analysis with explicit formulas of the variance matrix in special networks that includes 
 star networks and complete networks in \cite{WuXian2}. 
\par 
From Fig. \ref{fig.frequency}(c-d), it is seen
that as $\text{M}_t$ and $\text{D}_t$ increase,  $\|\bm f_\omega\|_\infty$ decreases respectively. This 
is consistent with the analysis in \cite{WANG2023110884} and \cite{WuXian2} 
on the dependence of $\bm\sigma_\omega^2$ on the inertia and damping coefficients.
\par 
Comparing the figures of $\overline{t}_e$ and $\|\bm f_\omega\|_\infty$ in Fig.\ref{fig.frequency}(a-d)
 and (e-h), we find that the trend of $\|\bm f_\omega\|_\infty$ fully reflects the 
 dependence of $\overline{t}_e$ on the system parameters. Thus, CREP characterizes
 the mean first hitting time consequently assesses the transient stability of power systems.

\subsubsection{Performance of minimizing $\Phi_\omega$} 
\begin{figure}
    \centering 
    \includegraphics[scale=1.05]{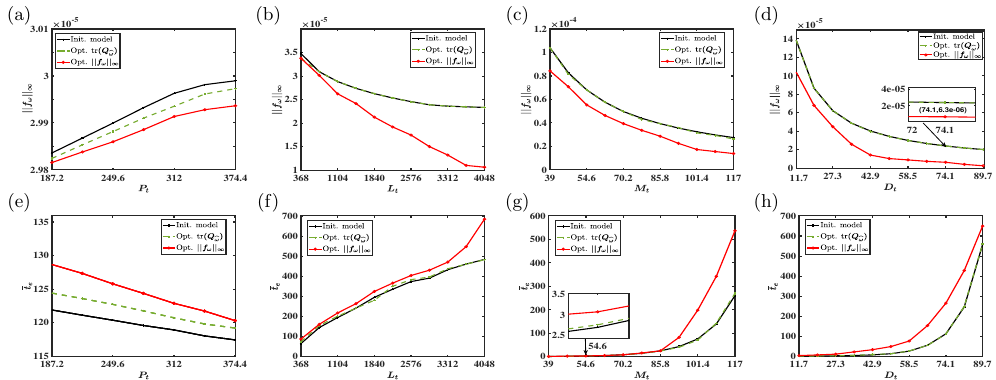}
    \caption{The dependence of $\overline{t}_e$ and $\|\bm f_\omega\|_\infty$ 
    on the parameters $\text{P}_t$, $\text{L}_t$, $\text{M}_t$ and $\text{D}_t$ in the models with the 3 configurations of paremeters.    }\label{fig.frequency}
\end{figure}

By comparing the curves of $\overline{t}_e$ in Fig. \ref{fig.frequency}(e-h),
the mean first hitting time when $\|\bm f_\omega\|_\infty$ is minimized 
is the largest one among the three metrics.
This demonstrates that the proposed optimization framework is the most effective on increasing the transient stability.

In contrast, it is surprising found from Fig. \ref{fig.frequency}(c-d) and (g-h) that 
the curves of $\|\bm f_\omega\|_\infty$ and $\overline{t}_e$
in the initial model and the model where $\text{tr}(\bm Q_{\widehat{\omega}})$ 
are minimized, almost overlap. This indicates 
that by minimizing $\text{tr}(\bm Q_{\widehat{\omega}})$ 
with either the inertia $M_i$ or the damping $D_i$ as decision variables,
the stability can hardly be improved.

\newpage
\onecolumn 
 \begin{table*}
        \begin{center}
            \caption{Configuration of the system parameters of evaluating the performance of $\|\bm f_\delta\|_\infty$.}
            \label{Parameters}
                \begin{tabular}{|c|c|c|c|c|c|c|c|}
                  \hline
                  ~$P_{2i-1},i=1,\cdots,20$~&~$P_2$~&~$P_{2i},i=2,\cdots, 19$~&~$l_{i,j}, (i,j)\in\mathcal{E}$~&~ $m_i$~&~$d_i$\\ \hline
 $-4$ & $8$ & $4$ & $20$ & $0.08i$ & $0.2\times(42-i)$
 \\ \hline 
            \end{tabular}
        \end{center}
      \end{table*}
\par 
      \begin{table*}
          \begin{center}
              \caption{Comparison of $\|\bm f_\delta\|$ with $\min\{|\text{Re}\mu_i|\}$ and the order parameters on identifying the Braess' paradox. The confidence intervals of $\overline{t}_e$ with confidence level $95\%$ are $[\overline{t}_e-t_c,\overline{t}_e+t_c]$ with $t_c\leq 2$s in all the 4 cases. }
              \label{paradox}
          \begin{tabular}{|c|c|cc|c|c|c|c|}
          \hline
          \multirow{2}{*}{Case} & \multirow{2}{*}{Added line} & \multicolumn{2}{c|}{Line capacity} &\multirow{2}{*}{$\overline{t}_e$} & \multirow{2}{*}{$||\bm f_{\bm\delta}||_{\infty}$} & \multirow{2}{*}{$\min{|\text{Re}(\mu_i)|}$} & \multirow{2}{*}{$\gamma$} \\ \cline{3-4}
           &  & \multicolumn{1}{c|}{$(22,35)$} & others & &  &  &  \\ \hline
          1 & $-$ & \multicolumn{1}{c|}{20} & 20 &195.14s&  $4.383\times 10^{-6}$& $0.2833$ &  0.9663\\ \hline	
          2 & $(19,23)$ & \multicolumn{1}{c|}{20} & 20 &148.94s &$1.419\times 10^{-5}$ & $0.3179$ & 0.9666\\ \hline		
          3 & $(24,38)$ & \multicolumn{1}{c|}{20} & 20 &241.51s &$3.393\times 10^{-6}$&  $0.2832$ &  0.9652 \\ \hline	
          4 & $-$ & \multicolumn{1}{c|}{30} & 20 &188.11s&$4.967\times 10^{-6}$ &   $0.3103$ &  0.9666 \\ \hline
          \end{tabular}
      \end{center}
          \end{table*}


\twocolumn

\section{Conclusion}\label{Section:conclusion}

Based on the theory of the invariant probability distribution of a stochastic process driven by 
Brownian motion, we have proposed a metric named CREP, which involves all the 
system parameters and reflects the size of the basin of attraction, 
to assess the transient stability. An optimization framework minimizing CREP
with the system parameters as decision variables was formulated.  
The mean first hitting time of the state hitting the boundary of a critical set, can be significantly increased by this approach, which
intuitively shows the strong potential of our approach in enhancing the transient stability.  
\par 
Future study will be on efficient algorithms for solving the corresponding optimization problems and theoretical analysis of the transient stability enhancement of power systems 
with non-Gaussian noise \cite{Schafer2018}. Extensions of the method for robustness improvement of the other nonlinear systems with continuously occurring disturbances will also be investigated. 

\appendix

\section{The invariant probability distribution and $\mathcal{H}_2$ norm}\label{InvariantDistribution}
Consider a linear time-invariant system, 
\begin{subequations}\label{Appendix:generalform}
 \begin{align}
  \dot{\bm x}&=\bm A\bm x+\bm B\bm v,\\
  \bm y&=\bm C\bm x, 
 \end{align}
\end{subequations}
where $\bm x\in\mathbb{R}^{n_x}$, $\bm A\in\mathbb{R}^{n_x\times n_x}$ is Hurwitz, $\bm B\in\mathbb{R}^{n_x\times n_v}$,  $\bm C\in\mathbb{R}^{n_y\times n_x}$, 
the input is denoted by $\bm v\in\mathbb{R}^{n_v}$ and the output of the system is denoted by $\bm y\in\mathbb{R}^{n_y}$. 
The squared $\mathcal{H}_2$ norm of the transfer matrix $\bm G$ of the mapping $(\bm A,\bm B,\bm C)$ from the input $\bm v$ to the output $\bm y$ is defined as 
\begin{subequations}\label{Appendix:H2norm}
 \begin{align}
  &||\bm G||^2_2=\text{tr}(\bm B^T\bm Q_o\bm B)=\text{tr}(\bm C\bm Q_c\bm C^T),\\
  &\bm Q_o\bm A+\bm A^T\bm Q_o+\bm C^T\bm C=\bm 0,\label{Lyapunov:equation}\\
  &\bm A\bm Q_c+\bm Q_c\bm A^T+\bm B\bm B^T=\bm 0,\label{Lyapunov:equation2}
 \end{align}
\end{subequations}
where $\text{tr}(\bm \cdot)$ denotes the trace of a matrix, $\bm Q_o,\bm Q_c\in\mathbb{R}^{n_x\times n_x}$ are the \emph{observability Grammian} of $(\bm C,\bm A)$ and \emph{controllability 
Grammian} of $(\bm A,\bm B)$ respectively \cite{H2norm_another_form,H2norm_book_toscano}. 
When the input $\bm v$ is modelled by Gaussian white noise,  the distribution of the state $\bm x$ and the output $\bm y$ are also Gaussian. 
Denote then for all $t \in T$,
$\bm x(t) \in G(\bm m_x(t), ~ \bm Q_{x}(t))$ with $\bm{Q}_{x}(t)\in \mathbb{R}^{n_x \times n_x}$ and 
$\bm y(t) \in G(\bm m_y(t), ~ \bm Q_{y}(t))$ with $\bm{Q}_{y}(t)\in \mathbb{R}^{n_y \times n_y}$.
Because the matrix $\bm A$ is Hurwitz, there exists an invariant probability distribution
of this linear stochastic system 
with the representation and properties
\begin{eqnarray*}
\mathbf{0}
    & = & \lim_{t \rightarrow \infty} ~ \mathbf{m}_x(t), ~
          \bm 0 = \lim_{t \rightarrow \infty} ~ \bm m_y(t), 
          \nonumber \\
        \bm Q_x 
    & = & \lim_{t \rightarrow \infty} ~ \bm Q_{x}(t), 
          \bm Q_y = \lim_{t \rightarrow \infty} ~ \bm Q_{y}(t), 
\end{eqnarray*}
where the variance matrices are
\begin{eqnarray*}
\bm Q_x
   & = & \int_0^{+\infty} \exp(\bm A t) 
          \bm B \bm B^{\top}
         \exp(\bm A^{\top} t) 
          \text{d}t,~~
          \bm Q_y=\bm C \bm Q_x \bm C^{\top}.
\end{eqnarray*}
Here $\bm Q_x$ is the unique solution of the Lyapunov matrix function (\ref{Lyapunov:equation2}) . 
\par 

\section{The  traditional optimization frameworks}\label{optimization-Case-study}

In this section, we present the traditional metric 
for the optimal configuration of the system parameters, the $\mathcal{H}_2$ norm of the 
system (\ref{linearization system}),  the phase cohesiveness and the order parameters for the level of the synchronization. 
\par 
The $\mathcal{H}_2$ norm of the system (\ref{linearization system}) where the term $\bm B\bm v(t)$ is seen as input to 
the system, 
is actually the trace of matrix $\bm Q_{\widehat{y}}$.  To minimize 
this norm, the optimization framework is 
\begin{align}
    &\min_{\bm \theta}~\text{tr}(\bm Q_{\widehat{\bm\delta}})\label{optimalNorm}\\
    \nonumber
    \text{s.t.} ~~&\text{(\ref{syn state}),~(\ref{output}),(\ref{decomposition}), (\ref{Qy}), (\ref{Qx})},
    (\ref{ConstaintPhase}),
    (\ref{decisionconstraint}). 
\end{align}
If the maximum of the variance of the phase angle differences in the edges is minimized, 
the objective function is replaced by $||\bm\sigma^2_\delta||_\infty$ in (\ref{optimalNorm}).  
The decision variables $\bm\theta$ can be either the power generation, the inertia, the damping 
coefficients or the line capacities and the corresponding constraints (\ref{decisionconstraint})
can be replaced by the ones in (\ref{OptimalGeneration}), (\ref{OptimalInertia}), (\ref{OptimalDamping})
and (\ref{OptimalCapacity}) respectively. 
\par 
The optimization framework for improving the phase cohesiveness is 
\begin{align}
    &\min_{\bm \theta}~\|\bm y_{\bm\delta}^*\|_\infty,\label{optimalinfinity}\\
    \nonumber
    \text{s.t.} ~~&\text{(\ref{syn state}),~(\ref{output}),(\ref{ConstaintPhase}),(\ref{decisionconstraint})}. 
\end{align}
\par
The order parameter of couple phase oscillators is defined 
as 
\begin{align}\label{orderparameter}
\gamma e^{\text{i}\phi}=\frac{1}{n}\sum_{j=1}^{n}e^{\text{i}\delta_j}
\end{align}
where $\text{i}^2=-1$ and $\delta_j$ 
is the phase at node $j$ and $\gamma e^{\text{i}\phi}$ is the phase' centroid on the 
complex unit circle with the magnitude $\gamma$ ranging from 0 to 1 \cite{kuramotobook}. 
In Section \ref{Section:Case study}, the order parameter is maximized by solving the following optimization problem \cite{Skardal2014},
\begin{align}
&\max_{\bm \theta}\gamma=1-||\bm\delta^*||_{2}^2/n,\label{optimalOrder}\\
\nonumber
&\text{s.t}~~\text{(\ref{syn state}),(\ref{ConstaintPhase}),(\ref{decisionconstraint})}. 
\end{align}
The decision variables $\bm\theta$ can be either the power generation or the line capacities and the corresponding constraints (\ref{decisionconstraint})
can be replaced by the ones in (\ref{OptimalGeneration})
and (\ref{OptimalCapacity}) respectively. 
In (\ref{optimalinfinity}) and (\ref{optimalOrder}), 
because the inertia and the damping of the synchronous machines have no impacts 
on the synchronous state, these parameters cannot be configured in an optimal way 
by these frameworks. 

{
\bibliographystyle{plain} 
\bibliography{autosam}

\begin{thebibliography}{10}

\bibitem{hirsch1}
H.~D. Chiang, M.W. Hirsch, and F.F. Wu.
\newblock {Stability regions of nonlinear autonomous dynamical systems}.
\newblock {\em IEEE Trans. Autom. Control}, 33(1):16--27, jan 1988.

\bibitem{Coletta2016}
T.~Coletta and P.~Jacquod.
\newblock {Linear stability and the Braess paradox in coupled-oscillator
  networks and electric power grids}.
\newblock {\em Phys. Rev. E}, 93(3):032222, mar 2016.

\bibitem{sizeofbasin}
R.~Delabays, M.~Tyloo, and Ph. Jacquod.
\newblock The size of the sync basin revisited.
\newblock {\em Chaos}, 27(10):103109, 2017.

\bibitem{DorflerCriticalcoupling}
F.~D{\"{o}}rfler and F.~Bullo.
\newblock {On the critical coupling for Kuramoto oscillators}.
\newblock {\em SIAM J. Appl. Dyn. Syst.}, 10(3):1070--1099, 2011.

\bibitem{Dorfler20141539}
F.~D{\"{o}}rfler and F.~Bullo.
\newblock Synchronization in complex networks of phase oscillators: A survey.
\newblock {\em Automatica}, 50(6):1539 -- 1564, 2014.

\bibitem{H2norm_another_form}
J.~C. Doyle, K.~Glover, P.~P. Khargonekar, and B.~A. Francis.
\newblock State-space solutions to standard {H}2 and {H} infinty control
  problems.
\newblock {\em IEEE Trans. Autom. Control}, 34(8):831--847, Aug 1989.

\bibitem{FAZLYAB2017181}
M.~Fazlyab, F.~D{\"{o}}rfler, and V.~M. Preciado.
\newblock {Optimal network design for synchronization of coupled oscillators}.
\newblock {\em Automatica}, 84:181 -- 189, 2017.

\bibitem{stochasticstability}
M.~M. Klosek-Dygas, B.~J. Matkowsky, and Z.~Schuss.
\newblock Stochastic stability on nonlinear oscillators.
\newblock {\em SIAM Journal on Applied Mathematics}, 48(5):1115--1127, 1988.

\bibitem{Kundur1994}
P.~Kundur.
\newblock {\em {Power system stability and control}}.
\newblock McGraw-Hill, 1994.

\bibitem{kuramotobook}
Y.~Kuramoto.
\newblock {\em Chemical oscillations, waves and turbulence}.
\newblock Springer, New York, 1984.

\bibitem{survivalanalysis}
M.~T. Lee and G.~A. Whitmore.
\newblock Threshold regression for survival analysis: Modeling event times by a
  stochastic process reaching a boundary.
\newblock {\em Statistical Science}, 21(4):501--513, 2006.

\bibitem{menck2}
P.~J Menck, J.~Heitzig, J.~Kurths, and H.~{Joachim Schellnhuber}.
\newblock {How dead ends undermine power grid stability}.
\newblock {\em Nat. Commun.}, 5:3969, jun 2014.

\bibitem{menck}
P.~J. Menck, J.~Heitzig, N.~Marwan, and J{\"{u}}rgen Kurths.
\newblock {How basin stability complements the linear-stability paradigm}.
\newblock {\em Nat. Phys.}, 9(2):89--92, jan 2013.

\bibitem{pecora}
L.~M. Pecora and T.~L. Carroll.
\newblock Master stability functions for synchronized coupled systems.
\newblock {\em Phys. Rev. Lett.}, 80:2109--2112, Mar 1998.

\bibitem{optimal_inertia_placement}
B.~K. Poolla, S.~Bolognani, and F.~D{\"{o}}rfler.
\newblock {Optimal placement of virtual inertia in power grids}.
\newblock {\em IEEE Trans. Autom. Control}, 62(12):6209--6220, 2017.

\bibitem{Schafer2018}
B.~Sch{\"{a}}fer, C.~Beck, K.~Aihara, D.~Witthaut, and M.~Timme.
\newblock {Non-Gaussian power grid frequency fluctuations characterized by
  L{\'{e}}vy-stable laws and superstatistics}.
\newblock {\em Nature Energy}, 3(2):119--126, 2018.

\bibitem{skar_uniqueness_equilibrium}
S.~J. Skar.
\newblock Stability of multi-machine power systems with nontrivial transfer
  conductances.
\newblock {\em SIAM J. Appl. Math.}, 39(3):475--491, 1980.

\bibitem{Skardal2014}
P.~S. Skardal, D.~Taylor, and J.~Sun.
\newblock Optimal synchronization of complex networks.
\newblock {\em Phys. Rev. Lett.}, 113:144101, Sep 2014.

\bibitem{H2norm}
E.~Tegling, B.~Bamieh, and D.~F. Gayme.
\newblock The price of synchrony: Evaluating the resistive losses in
  synchronizing power networks.
\newblock {\em IEEE Trans. Control Netw. Syst.}, 2(3):254--266, Sept 2015.

\bibitem{H2norm_book_toscano}
R.~Toscano.
\newblock {\em Structured controllers for uncertain systems}.
\newblock Springer-verlag, London, 2013.

\bibitem{WANG2023110884}
Z.~Wang, K.~Xi, A.~Cheng, H.~X. Lin, A.~C.M. Ran, J.~H. {van Schuppen}, and
  C.~Zhang.
\newblock Synchronization of power systems under stochastic disturbances.
\newblock {\em Automatica}, 151:110884, 2023.

\bibitem{Witthaut2012}
D.~Witthaut and M.~Timme.
\newblock {Braess's paradox in oscillator networks, desynchronization and power
  outage}.
\newblock {\em New J. Phys.}, 14(8):083036, aug 2012.

\bibitem{WuXian}
X.~Wu, K.~Xi, A.~Cheng, H.~X. Lin, and J.~H. van Schuppen.
\newblock {Increasing the synchronization stability in complex networks}.
\newblock {\em Chaos: An Interdisciplinary Journal of Nonlinear Science},
  33(4), 04 2023.
\newblock 043116.

\bibitem{WuXian2}
X.~Wu, K.~Xi, A.~Cheng, H.~X. Lin, J.~H. van Schuppen, and C.~Zhang.
\newblock {Explicit formulas for the variance of the state of a linearied power
  system driven by Gaussian stochastic disturbances}.
\newblock {\em preprint in arXiv:2302.06326}, 2023.

\bibitem{Xi2016}
K.~Xi, J.~L.~A. Dubbeldam, and H.~X. Lin.
\newblock {Synchronization of cyclic power grids: equilibria and stability of
  the synchronous state}.
\newblock {\em Chaos}, 27(1):013109, 2017.

\bibitem{PIAC3}
K.~Xi, J.~L.A. Dubbeldam, H.~X. Lin, and J.~H. van Schuppen.
\newblock {Power-Imbalance Allocation Control of Power Systems-Secondary
  Frequency Control}.
\newblock {\em Automatica}, 92:72 -- 85, 2018.

\bibitem{zab1}
J.~Zaborsky, G.~Huang, T.~C. Leung, and B.~Zheng.
\newblock {Stability monitoring on the large electric power system}.
\newblock In {\em 24th IEEE Conf. Decision Control}, volume~24, pages 787--798.
  IEEE, dec 1985.

\bibitem{zab2}
J.~Zaborszky, G.~Huang, B.~Zheng, and T.~C. Leung.
\newblock {On the phase portrait of a class of large nonlinear dynamic systems
  such as the power system}.
\newblock {\em IEEE Trans. Autom. Control}, 33(1):4--15, jan 1988.

\bibitem{Zha19}
X.~Zhang, S.~Hallerberg, M.~Matthiae, D.~Witthaut, and M.~Timme.
\newblock Fluctuation-induced distributed resonances in oscillatory networks.
\newblock {\em Sci. Adv.}, 5(7):eaav1027, 2019.

\bibitem{topological-spreading}
X.~Zhang, D.~Witthaut, and M.~Timme.
\newblock Topological determinants of perturbation spreading in networks.
\newblock {\em Phys. Rev. Lett.}, 125:218301, 2020.

\end{thebibliography}
}

\end{document}